\documentclass[nonacm]{acmart}
\usepackage{subfig}
\usepackage{tabularx}
\usepackage{xcolor}
\usepackage{soul}
\usepackage{placeins}
\usepackage{url}
\usepackage{hyperref}
\AtBeginDocument{%
  \providecommand\BibTeX{{%
    \normalfont B\kern-0.5em{\scshape i\kern-0.25em b}\kern-0.8em\TeX}}}

\setcopyright{acmcopyright}
\copyrightyear{2022}
\acmYear{2022}
\acmDOI{XXXXXXX.XXXXXXX}

\acmJournal{TOG}
\acmVolume{37}
\acmNumber{4}
\acmArticle{111}
\acmMonth{8}

\begin{document}

\title{A DNA Based Colour Image Encryption Scheme Using A Convolutional Autoencoder}

\author{Fawad Ahmed}
\email{fawad@pnec.nust.edu.pk}
\affiliation{%
\department{Department of Cyber Security}
  \institution{Pakistan Navy Engineering College, NUST}
  \city{Karachi}
  \country{Pakistan}
  \postcode{75350}
}

\author{Muneeb Ur Rehman}
  \email{muneeb95rehman@gmail.com}
\affiliation{%
\department{Department of Electrical Engineering}
  \institution{HITEC University}
  \city{Taxila}
  \country{Pakistan}
  }

\author{Jawad Ahmad}
\affiliation{%
  \institution{School of Computing, Edinburgh Napier University}
  \city{Edinburgh}
  \country{United Kingdom}}

\author{Muhammad Shahbaz Khan}
\email{shahbaz.khan@hitecuni.edu.pk}
\affiliation{%
    \department{Department of Electrical Engineering}
  \institution{HITEC University}
  \city{Taxila}
  \country{Pakistan}
  \postcode{47080}}

\author{Wadii Boulila}
\email{wboulila@psu.edu.sa}
\affiliation{%
\department{Robotics and Internet-of-Things Laboratory}
  \institution{Prince Sultan University}
  \city{Riyadh}
  \country{Saudi Arabia}}
\affiliation{
 \department{RIADI Laboratory}
  \institution{University of Manouba}
  \city{Manouba}
  \country{Tunisia}
}
  
   \author{Gautam Srivastava}
   \email{srivastavag@brandonu.ca}
\affiliation{%
    \department{Math and Computer Science}
  \institution{Brandon University}
  \city{Brandon}
  \country{Canada}}
\affiliation{
 \department{Research Centre for Interneural Computing}
  \institution{China Medical University}
  \city{Taichung}
  \country{Taiwan}
}
\affiliation{
 \department{Dept of Computer Science and Math}
  \institution{Lebanese American University}
  \city{Beirut}
  \country{Lebanon}
}
  
  \author{Jerry Chun-Wei Lin}
  \email{jerrylin@ieee.org}
  \affiliation{%
  \institution{Western Norway University of Applied Sciences}
  \city{Bergen}
  \country{Norway}}

  \author{William J. Buchanan}
  \email{B.Buchanan@napier.ac.uk}
\affiliation{%
  \institution{School of Computing, Edinburgh Napier University}
  \city{Edinburgh}
  \country{United Kingdom}}
\renewcommand{\shortauthors}{Fawad Ahmed, et al.}

\begin{abstract}
With the advancement in technology, digital images can easily be transmitted and stored over the Internet. Encryption is used to avoid illegal interception of digital images. Encrypting large-sized colour images in their original dimension generally results in low encryption/decryption speed along with exerting a burden on the limited bandwidth of the transmission channel. To address the aforementioned issues, a new encryption scheme for colour images employing convolutional autoencoder, DNA and chaos is presented in this paper. The proposed scheme has two main modules, the dimensionality conversion module using the proposed convolutional autoencoder, and the encryption/decryption module using DNA and chaos. The dimension of the input colour image is first reduced from $N\times M\times 3$ to $P \times Q$ gray-scale image using the encoder. Encryption and decryption are then performed in the reduced dimension space. The decrypted gray-scale image is upsampled to obtain the original colour image having dimension $N \times M \times 3$. The training and validation accuracy of the proposed autoencoder is 97\% and 95\%, respectively. Once the autoencoder is trained, it can be used to reduce and subsequently increase the dimension of any arbitrary input colour image. The efficacy of the designed autoencoder has been demonstrated by the successful reconstruction of the compressed image into the original colour image with negligible perceptual distortion. The second major contribution presented in this paper is an image encryption scheme using DNA along with multiple chaotic sequences and substitution boxes. The security of the proposed image encryption algorithm has been gauged using several evaluation parameters, such as histogram of the cipher image, entropy, NPCR, UACI, key sensitivity, contrast, etc. The experimental results of the proposed scheme demonstrate its effectiveness to perform colour image encryption.

\end{abstract}




\maketitle
\pagestyle{plain}
\section{Introduction}
Due to advancements in network communication, the Internet of Things (IoT), telemedicine, online biometric systems, and social media, a large amount of digital images are transmitted over the Internet. The information contained in these digital images can be illegally intercepted and tampered during the transmission or storage process. Hence, there is an indispensable need to secure the transmission and storage of digital images, and in this context, image encryption plays a vital role  \cite{kumar2021secured}. Various encryption schemes, such as, DES, 3DES, AES, RSA, etc., are {most commonly} used for {securing digital} data over open networks. But it is pertinent to mention {here} that {the} data pattern in images is {entirely different from the text data.} Images contain a large number of pixels {, and they have a high degree of correlation between their adjacent pixels} \cite{lai2022hidden}. {Due to this reason,} securing image data is more complex than the text data and 
 {hence,} the aforementioned text-based encryption schemes are often unsuitable for image encryption \cite{shafique2020detecting}. Moreover, {for the high definition (HD) colored images}, these conventional encryption schemes involve extra operations, which result in increased computational time and power. Consequently, these schemes exhibit low encryption and decryption speeds that may lead to significant delays in real-time communication. In recent years, there has been a surge in the number of image encryption schemes proposed in the literature. The application of chaos {theory} and DNA sequencing in image encryption has proved to be {quite} effective and efficient in securing digital images  \cite{pak2017new,wang2020hyperchaotic,chai2017image,zhang2018image}. {Therefore, the main motivation of this paper was to develop a chaos and DNA-based image encryption scheme, specifically for HD-colored images that should not only be highly secure but should also address the latency issues faced during the encryption and transmission processes.}

 {The} chaotic maps possess several interesting properties such as pseudo-randomness, unpredictability, mixing (ergodicity), and high sensitivity due to change in control parameters and initial conditions  \cite{lai2022hidden,masood2022novel,chai2017image,zhang2018image}. {In addition to these properties}, chaotic maps also have reduced mathematical complexity and a {higher} level of security. DNA computing, {on the other hand,} involves {utilization of DNA} coding sequences for carrying {and securing digital} information  \cite{liu2012rgb}. These coding sequences are generated {by} using four types of DNA nucleotide bases, i.e., A, C, G and T. DNA computing performs exceptionally well when dealing with large data and parallelism, resulting in reduced {computational power during the} transmission of digital information. DNA coding in conjunction with chaotic maps has opened a new paradigm of image encryption  \cite{xue2010digital,masood2021new,xue2020new} {and hence, needs to be properly explored to propose more secure and highly efficient encryption schemes.}  

\indent As described previously, {the} image data is quite large as compared to text data. {Therefore,} storing and transmitting such a large amount of data (especially HD {colored} images) is quite challenging. {In addition to} increasing the computational complexity of the encryption algorithm, it also increases the bandwidth requirement of {the} transmission channel as well. This makes it difficult to apply complex image encryption algorithms on large-size {colored} images {in real-time. For better encryption and decryption speeds,} recent literature reveals that the deep neural networks based autoencoders \cite{hu2017batch} exhibit promising results. Such schemes are based on the encoder-decoder paradigm for dimensionality reduction, which helps to decrease the computational complexity of the encryption algorithm. {Autoencoders solve the latency issues by} first reducing the dimension of the input colour image and then applying the encryption algorithm in the reduced dimension space. Due to less number of pixels in the reduced dimension space, it becomes comparatively easy to practically implement complex image encryption algorithms {on edge devices. Thus, the motivation to use an autoencoder-based encryption scheme is to achieve real-time secure image transmission/storage without exerting a burden on the available bandwidth, especially on edge devices in IoT \cite{latif2021deep}.}

\begin{figure}[t]
\captionsetup{justification=centering}
\centering\includegraphics[scale=1]{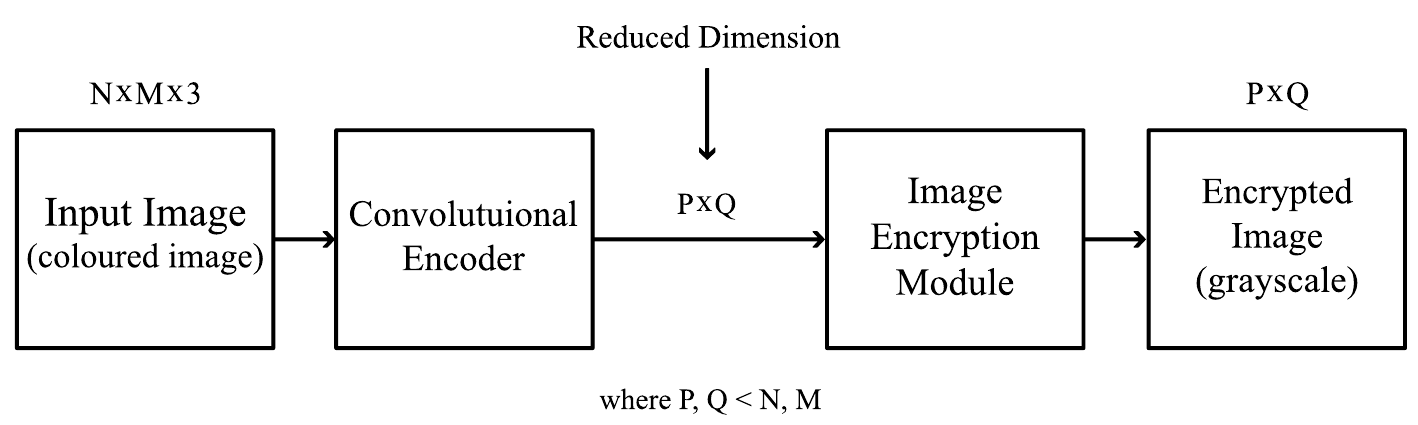}
\caption{Basic Workflow of Image Encryption.}
\label{block1}
\end{figure}
\begin{figure}[t]
\captionsetup{justification=centering}
\centering\includegraphics[scale=.96]{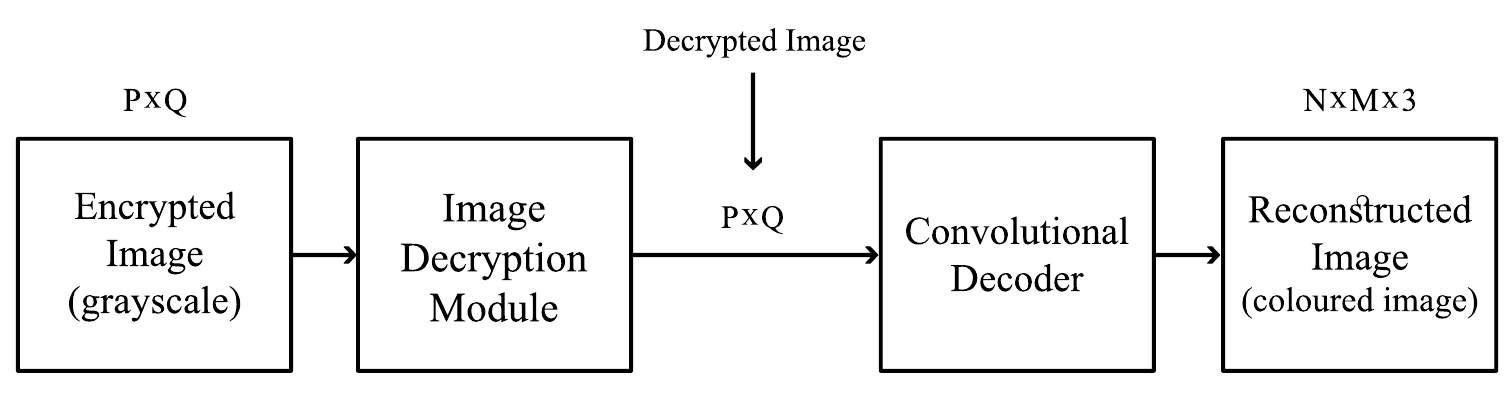}
\caption{Basic Workflow of Image Decryption.}
\label{block2}
\end{figure}
\indent {Various DNA-based encryption schemes can be found in the literature that exhibits slow encryption and decryption schemes. Despite being secure enough, these schemes cannot be implemented in real-time on today’s small-size IoT devices. For instance, the first method to use DNA computing for data encryption was the Hiding Messages in Microdots \cite{new1} scheme, but it was extremely slow. Even some of the recent and novel techniques, such as the Public Key System by using DNA (PKSDNA) \cite{new2}, the Chaos-Based Image Encryption (CBIE) \cite{new3}, the DNA-based Reversible Data Hiding Scheme (RDHS) \cite{new4}, and the Double-Layer Data Hiding scheme (DLDH) \cite{new5} pose latency issues. These schemes take a lot of time for encrypting and decrypting multimedia files, and also while generating or retrieve secret keys.}
\\
\indent {To solve the aforementioned problems of existing schemes}, this paper presents a new autoencoder based {image encryption} scheme for large size {colored} images. The proposed scheme consists of two modules; 1) A deep learning-based convolutional autoencoder for dimensionality reduction module, and 2) A Chaos and DNA-based image encryption module.{The basic workflow of the proposed image encryption and decryption modules is given in Fig. \ref{block1} and Fig. \ref{block2}, respectively.} A colour image having a dimension of  N$\times$M$\times$3 is first compressed to a lower dimension P$\times$Q using the proposed convolutional autoencoder model. Encryption is performed in the compressed domain. The encrypted image is then decrypted using the decryption module to retrieve back the plaintext image of dimension $P \times Q$. This decrypted image is then given to the decoder to get back the original colour image of dimension $N \times M \times 3$. It is pertinent to mention that the proposed autoencoder is designed to reconstruct the original colour image with negligible perceptual distortion as evident from the results presented in the paper. \\ 

\vspace{2mm}

\noindent {Following are the important contributions of this work:}

\begin{itemize}
   \item A convolution autoencoder is proposed to reduce the dimension of the input colour image. 
    
    \item A new image encryption technique is proposed that uses DNA, chaos, and multiple s-boxes to effectively encrypt and decrypt images in lower dimension space. 
    
    \item Due to the utilization of multi-chaotic maps, the proposed scheme has a significantly higher keyspace when compared with other schemes. 
    
   % \item Due to the obtained lower-dimensional space, the computational complexity, bandwidth, and storage space required by the encryption and decryption algorithms are reduced.

\end{itemize}

\indent The remaining sections are structured as follows. Recent literature covering image encryption schemes that use chaos, chaos+DNA, and deep learning is presented in Section \ref{sec2}. Important details related to the chaotic maps used in this paper are discussed in Section \ref{sec3}. In Section \ref{sec4}, the proposed image encryption scheme is presented. Section \ref{sec5} contains experimental results depicting the quality of reconstructed images and the encryption algorithm, thus proving the effectiveness of the proposed scheme. Finally, Section \ref{sec6} presents conclusion and future directions.

\section{Related Work}\label{sec2}
In the literature, several encryption schemes for images employing chaotic maps have been proposed \cite{g1,g2,g3}. Chaotic maps are used due to their inherent properties, such as unpredictability, pseudo-randomness, high sensitivity for control parameters and initial conditions, and mainly ergodicity. For example, various architectures using chaos to permute and diffuse digital images are presented in \cite{hoang2020novel}. Similarly for video encryption, a chaos-based scheme is presented in \cite{abu2020privacy} that uses the inherent property of the HEVC standard to achieve improved security and exhibit secure real-time video encryption with an optimal bit rate. In addition, chaos with multi-dimensional feature vectors has also been utilized in improving the security of steganographic approaches \cite{battikh2019comparative}. The results show that irrespective of the size of the message, the presented steganalysis system can effectively detect hidden information. Furthermore, a neural network based on chaos having key dependent hash functions is presented in \cite{abdoun2020designing}. Pseudo-chaotic samples have been generated that are fed into the neural structure as parameter values. The presented architecture helps in generating hash functions having good statistical features along with high message and key sensitivity. Moreover, the effectiveness of chaotic functions against cryptanalysis has been analyzed in \cite{dimitrov2020design}. For this purpose, various s-boxes have been designed by using two heuristic methods. The designed s-boxes exhibit exceptional results when compared to other s-boxes available in the literature.\\
\indent Moreover, in the last few years, chaos-based encryption has been utilized in conjunction with DNA computing to devise new image encryption schemes \cite{chai2017image,zhang2018image,zhen2016chaos}. DNA computing is the application of various biological and algebraic processes applied to DNA sequences, e.g., addition, XOR and subtraction operations on DNA coded sequences. These operations are recently being preferred to be used with chaos-based encryption schemes, for example, an image encryption algorithm based on a compound chaotic map and varying DNA coding is presented in \cite{zhang2021novel}. The proposed algorithm executes effectively in terms of security performance and can be compared with already available chaos-based encryption schemes. Similarly, another image encryption algorithm is presented in \cite{wang2021chaotic} that combines a 1D logistic map, DNA coding, and multi-objective particle swarm optimization (MOPSO). The presented results exhibit resistance of the proposed algorithm against several attacks along with excellent entropy and correlation-coefficient of the ciphertext. Apart from 1D chaotic systems, a 3D chaotic system has also been used with DNA coding in \cite{kengnou2021image}. The presented encryption scheme utilizes all 24 rules of DNA coding with 16 joint operations that make this scheme resistant to several attacks and an excellent scheme for securing digital images. Furthermore, DNA encryption and chaos have also been utilized for the encryption of coloured images \cite{chai2017novel}. This scheme utilizes permutation by shuffling the RGB components of the input colour image. The permuted components are then recombined by DNA encoding, and finally, diffusion is used on the encoded DNA matrix. {DNA encryption schemes are also being preferred for cloud environments \cite{new6}, \cite{new7}, \cite{new8}. In \cite{new6} DNA computing is used for generating a 1024-bit secret key, and a DNA reference key for better security. Similarly, in \cite{new7} for a key generation the DNA computing has been combined with user attributes and the MAC address of the user for improved security. DNA computing-based multifold symmetric key encryption scheme has been presented in \cite{new8} which encrypts data before uploading it to the cloud. DNA computing also plays a potential role in securing medical images in smart healthcare systems \cite{new9,new10,new11, new12}.}  

\indent In addition to chaos and DNA encryption, deep learning has recently been playing a potential role in image encryption. Most of the presented techniques utilize the encoder-decoder approach. Autoencoders are extensively being utilized for dimensionality reduction and to increase the networks’ performance. Recently, various stacked autoencoders and convolutional autoencoders have been explored and designed for encryption/decryption purposes, for example, a stacked autoencoder based on a multi-layer model for compression and encryption of images has been presented in \cite{hu2016image}. The first layer of the model compresses the input image. The compressed image is then encrypted using a chaotic logistic map. Such models hold importance in applications where transmission speed is also as important as the security of the images. Similarly, another framework utilizes a neural network-based encoder and decoder for encryption and decryption, respectively \cite{gupta2020shallow}. The presented framework is targeted for lossy encryption and decryption and exhibits adequate performance where exact reconstruction of the input image is not a strict requirement.\\

\indent Majority of the image encryption schemes available in the literature, including the ones discussed above have latency problems in case encryption and decryption of large-size images are required. This problem becomes more pronounced when the input image is a colour image. In such a case, there are three colour planes, thus increasing the size of the images by three times. Although several autoencoders can be found in literature, however, the reconstructed image quality is not very high. Moreover, most of the autoencoders available in the literature have been designed for low to medium size digital images. To address the aforementioned problem, the proposed scheme uses a new autoencoder design to effectively reduce the dimension of the input image from $N\times M \times 3$ to $P \times Q$. Thus a colour image is converted by the autoencoder into a grayscale image. For example, in this paper, the input colour images used have dimensions $512 \times 768 \times 3$ pixels. The autoencoder converts the input colour image to a gray-scale image having dimension $384 \times 512$ pixels. Encryption and decryption are then performed in the reduced dimension space using a new DNA-based technique employing multiple chaotic sequences and substitution boxes to make the cipher image random, as evident from the results presented in the paper.
%%%%%%%%%%%%%%%%%%%%%%%%%%%%%%%%%%%%%%%%%%
\section{Preliminaries}\label{sec3}
\subsection{Utilized Chaotic Maps}
\indent{There are two broad categories of chaotic systems: one-dimensional and multi-dimensional. Unlike multi-dimensional, one-dimensional chaotic systems such as logistic maps etc., are straightforward to implement. However, there are drawbacks of one-dimensional maps that include their susceptibility to high correlation, as well as their restricted or discontinuous chaotic range and high probability of uniform data distribution in the output chaotic sequence. Thus, in this work, we have used four multi-dimensional maps due to low correlation, higher keyspace and non-uniform data distribution.} 

Four different chaotic maps are used in this work to introduce randomness at different stages of the encryption process. The four chaotic maps are the TD-ERCS map, the Intertwining map, the Chirikov map, and the NCA map. A brief description of these maps is as follows:
\\
\\
\textbf{TD-ERCS Map}
\\
The TD-ERCS map is a two-dimension chaotic system proposed in the year 2004 and further developed to exhibit a large domain, zero correlation and a stable probability distribution \cite{liao2011td}. In this work, the TD-ERCS map is used for random permutation of pixel values. 
The TD-ERCS map is expressed as:

%%%%%%%%%%%%%%%%%%%%%%%%%%%%%%%%%
\begin{equation} \label{TD-ERCS2}
\mbox{} % if not included error it was c = any alphabet.
\begin{cases}
x_{n}= -\frac{2k_{n-1}y_{n-1}+x_{n-1}(\mu^2-k^2_{n-1})}{\mu^2+k^2_{n-1}} & \\
y_{n}= k_{n-1}(x_{n}-x_{n-1})+y_{n-1}, & n= 1,2,3 ... \\
\end{cases}
\end{equation}
%%%%%%%%%%%%%%%%%%%%%%%%%%%%%%%%
where\\
%%%%%%%%%%%%%%%%%%%%%%%%%%%%
\begin{displaymath}
k_{n} = \frac{2k{'_{n-m}}-k_{n-1}+k_{n-1}(k{'_{n-m}})^2}{1+2k_{n-1}k{'_{n-m}}-k(k{'_{n-m}})^2}
\end{displaymath}
%%%%%%%%%%%%%%%%%%%%%%%%%%%%
\begin{displaymath}
k{'_{n-m}}=
\begin{cases}
-\frac{x_{n-1}}{y_{n-1}}\mu^2 & n < m\\
-\frac{x_{n-m}}{y_{n-m}}\mu^2 & n \geq m \\
\end{cases}
\end{displaymath}
%%%%%%%%%%%%%%%%%%%%%%%%%%%%%%%%
\begin{displaymath}
y_{0}= \mu \sqrt{1-x_{0}^{2}}
\end{displaymath}
\begin{displaymath}
k{'_{0}}=- \frac{x_{0}}{y_{0}}\mu^2
\end{displaymath}

\begin{displaymath}
k_{0} = - \frac{\mbox{tan}\alpha + k{'_{0}}}{1-k{'_{0}}\mbox{tan}\alpha }
\end{displaymath}
\begin{displaymath} \label{TD-ERCS}
\left\{ 
  \begin{array}{l} \mu \in (0,1)\\
  x_{0} \in [-1,1] \\ 
  \alpha \in (0, \pi) \\
  m = 2,3,4,5 ...
  							 \end{array} \right .		 
\end{displaymath}
\\In the above equation, $\mu, x_{0}, \alpha$ and $m$ are called the initial parameters. As outlined in our previous work \cite{ahmad2016secure}, the TD-ERCS map holds the property of being sensitive to the initial conditions and ergodicity and hence can be effectively used for image encryption.\\ 
\\

\noindent\textbf{Intertwining Map}
\\The intertwining map is used for random matrix generation and is written as \cite{ahmad2018secure22}: 

\begin{equation} \label{Cheb}
	\begin{cases}
	X_{n+1} = (\lambda \times A_1 \times Y_{n} \times (1-X_{n})+Z_{n})mod(1), & \\
	Y_{n+1} = (\frac{\lambda \times A_2 \times Y_{n} +Z_{n} } {1+(X_{n+1})^2})mod(1), &\\
	Z_{n+1} = (\lambda \times (X_{n+1} + Y_{n+1} + A_3) \times sin(Z_{n})mod(1). & \\
	\end{cases}
\end{equation}
where $X_{n}$, $Y_n$ and $ Z_n \in (0,1)$, $0\leq \lambda \leq 3.999$, $|A_1| > 33.5, |A_2| > 37.9, |A_3| > 35.7$. \\

\noindent\textbf{Chirikov Map}
\\The Chirikov map is used to randomly select the DNA coding rules. Its mathematical model is as follows \cite{durafe2022comparative12}: 
	\begin{align}
	A_{n+1}&=(A_n+B_n)\mbox{ mod}(N),\\
	B_{n+1}&=(A_n + \eta \times \mbox{Sin}(\frac{2\pi A_{n}}{N}))\mbox{ mod}(N),
	\end{align}		
where $N$ is the height/width of a square image, $\eta > 0$ and is used as a control parameter for the secret key, $A_n$ and $B_n$ are real numbers whose range is between 0 to $N$.\\
\\
\noindent\textbf{NCA Map}
\\The Nonlinear Chaotic Algorithm (NCA) is employed for randomly selecting one of the three S-boxes. These S-boxes are discussed in our previous work \cite{qayyum2020chaos}. The NCA map is expressed as \cite{ahmad2018secure22}:
\begin{equation} \label{NCA}
C_{n+1} = (1- \xi^{-4}) . \text{cot}(\frac{\chi}{1+ \xi}) . (1+ \frac{1}{\xi})^{\xi}. \text{tan}(\chi C_{n}) . (1-C_{n})^{\xi},
\end{equation}
where the seed parameters are:
%%%%%%%%%%%%%%%%%%%%%%%%%%%%%%%%%%%%%%%%%%%%%%%%%%%%%%%%%%
\begin{equation*} 
\left\{ 
\begin{array}{l l l l l l l l l l l}       C_{n} \in (0,1) \\
\chi \in (0,1.4] \\
\xi \in [5,43] \\
\text{or} \\
C_{n} \in (0,1) \\
\chi \in (1.4,1.5] \\
\xi \in [9,38] \\
\text{or} \\     
C_{n} \in (0,1) \\
\chi \in (1.5,1.57] \\
\xi \in [3,15] 
\end{array} \right .		 
\end{equation*}

\vspace{10mm}

\subsection{DNA Coding}
Adenine (A), cytosine (C), guanine (G) and thymine are the nucleic bases to generate different DNA sequences. The nucleic bases, ‘adenine and thymine ’ and ‘guanine and cytosine’ are complements of each other. In the proposed encryption scheme, the aforementioned DNA nucleic bases A, T, G and C are represented by the binary codes 00, 01, 10, and 11, respectively. Hence, 24 different coding sequences can be produced. However, the binary codes ‘00 and 11’, and ‘01 and 10’ are complements of each other and hence, only 8 out of 24 coding rules satisfy the complementary criterion for the four bases. The 8 rules for DNA coding are shown in Table 1 \cite{xue2020new}. Furthermore, each pixel of a gray-scale image is represented by a value between 0 and 255. The binary representation of a pixel value can be divided into 4 groups, each group made up of two bits. The two bits of each group can then be represented using the DNA basis, depending on the selected rule. As an example, a pixel having a value of ‘200’ can be represented in binary as ‘11001000’, and by following the R2 rule given in Table \ref{tab:dna code} \cite{xue2020new}, the DNA sequence for this pixel is ‘TACA’. DNA decoding can be performed similarly by following the appropriate rule.

\begin{table}[!ht]
\center
\caption{The Eight DNA rules with binary coding.}
\label{tab:dna code}       % Give a unique label
\scalebox{0.98}{
\begin{tabular*}{\textwidth}{c @{\extracolsep{\fill}}ccccccccc}
\hline
\textbf{Code} & \textbf{R1} & \textbf{R2}& \textbf{R3}& \textbf{R4}& \textbf{R5}& \textbf{R6}& \textbf{R7}& \textbf{R8}  \\
\hline
00 & [A] & [A] & [C] & [C] & [G] & [G] & [T] & [T] \\
01 & [C] & [G] & [A] & [T] & [A] & [T] & [C] & [G] \\
10 & [G] & [C] & [T] & [A] & [T] & [A] & [G] & [C] \\
11 & [T] & [T] & [G] & [G] & [C] & [C] & [A] & [A] \\
\hline
\end{tabular*}}
\end{table}
%Table3 from doc
\iffalse
\begin{table}[!ht]
\center
\caption{DNA Sequence XOR operations.}
\label{tab:dna code3}       % Give a unique label
\scalebox{1.2}{
\begin{tabular}{*{10}{cl}}
\hline
\centering
\textbf{XOR}& \textbf{A}& \textbf{C}& \textbf{G}& \textbf{T}  \\
\hline
A & A & C & G & T\\
C & C & A & T & G \\
G & G & T & A & C\\
T & T & G & C & A \\
\hline
\end{tabular}}
\end{table}
\fi
\section{The Proposed Methodology}\label{sec4}
\begin{figure}[!t]
\captionsetup{justification=centering}
\centering\includegraphics[scale=0.52]{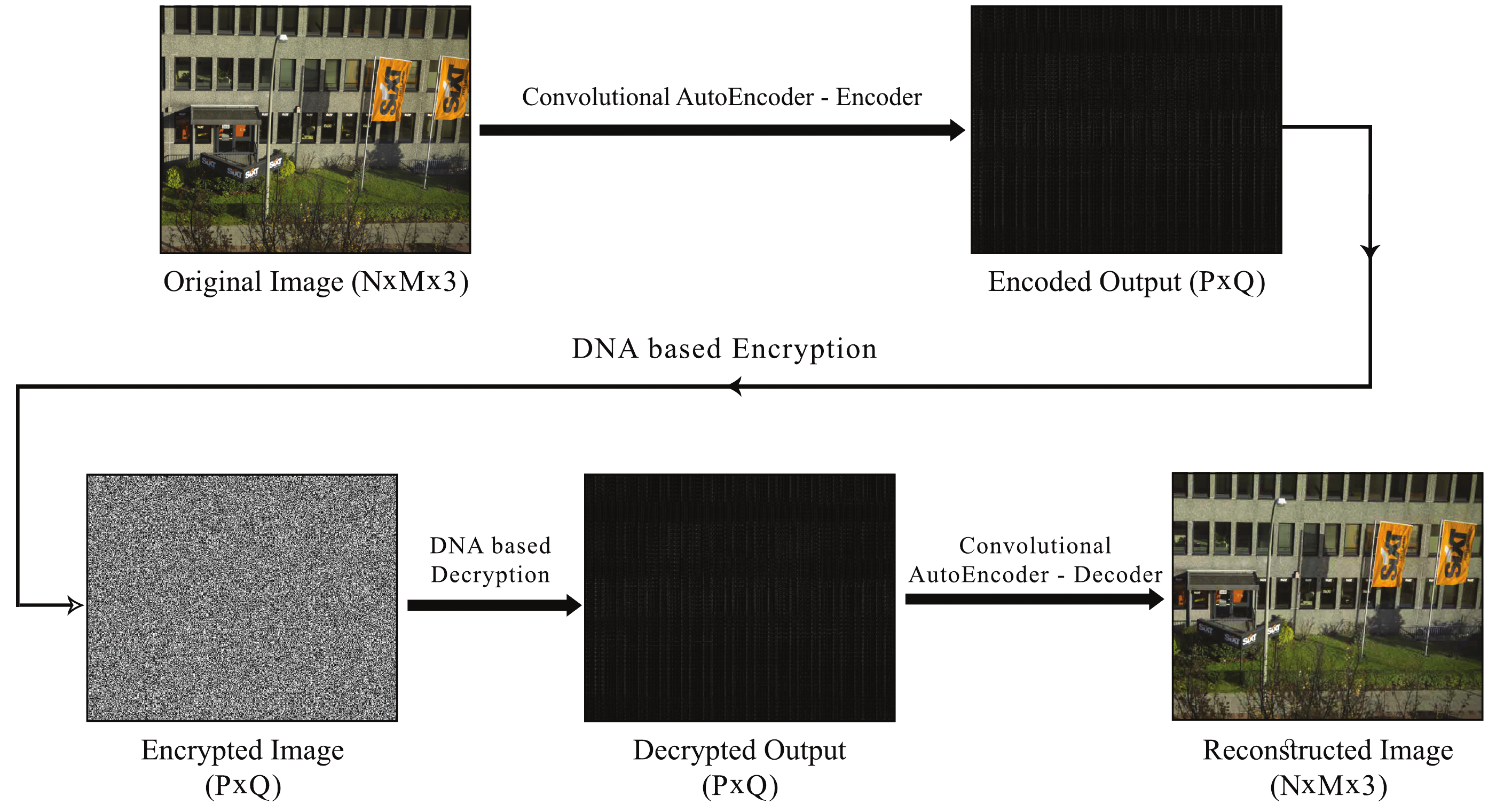}
\caption{A Pictorial Flow Depicting the Workflow of the Proposed Scheme.}
\label{schemeimage}
\end{figure}

The proposed scheme consists of two main parts, the Convolution Autoencoder-based dimensionality reduction module, and the DNA-based image encryption module. A colour input image of size $N \times M \times 3$ is first reduced to a lower dimension gray-scale image of size $P \times Q$ after passing through a trained convolution-based encoder model. The cipher image is obtained by applying the encryption algorithm on the $P \times Q$ gray-scale image. The encrypted image is then decrypted using the decryption module and subsequently decompressed using upsampling by the decoder to retrieve the original colour image with negligible loss of information. Figure ~\ref{schemeimage} shows the workflow of the proposed colour image encryption scheme and  Fig.~\ref{proposedschemedetail} shows the detailed architecture of the proposed autoencoder illustrating the encoder, decoder along with the encryption and decryption modules. {The Kodak dataset\footnote{\url{https://r0k.us/graphics/kodak/}} has been used in this paper for both the dimensionality reduction module and the encryption module.} The next section presents details about the proposed autoencoder module for dimensionality reduction.\\
\\
\subsection{The Proposed Convolutional Autoencoder for Dimensionality Reduction}
\begin{figure}[!ht]
  \centering
    \captionsetup{justification=centering}
    \subfloat[]{\includegraphics[scale=.45]{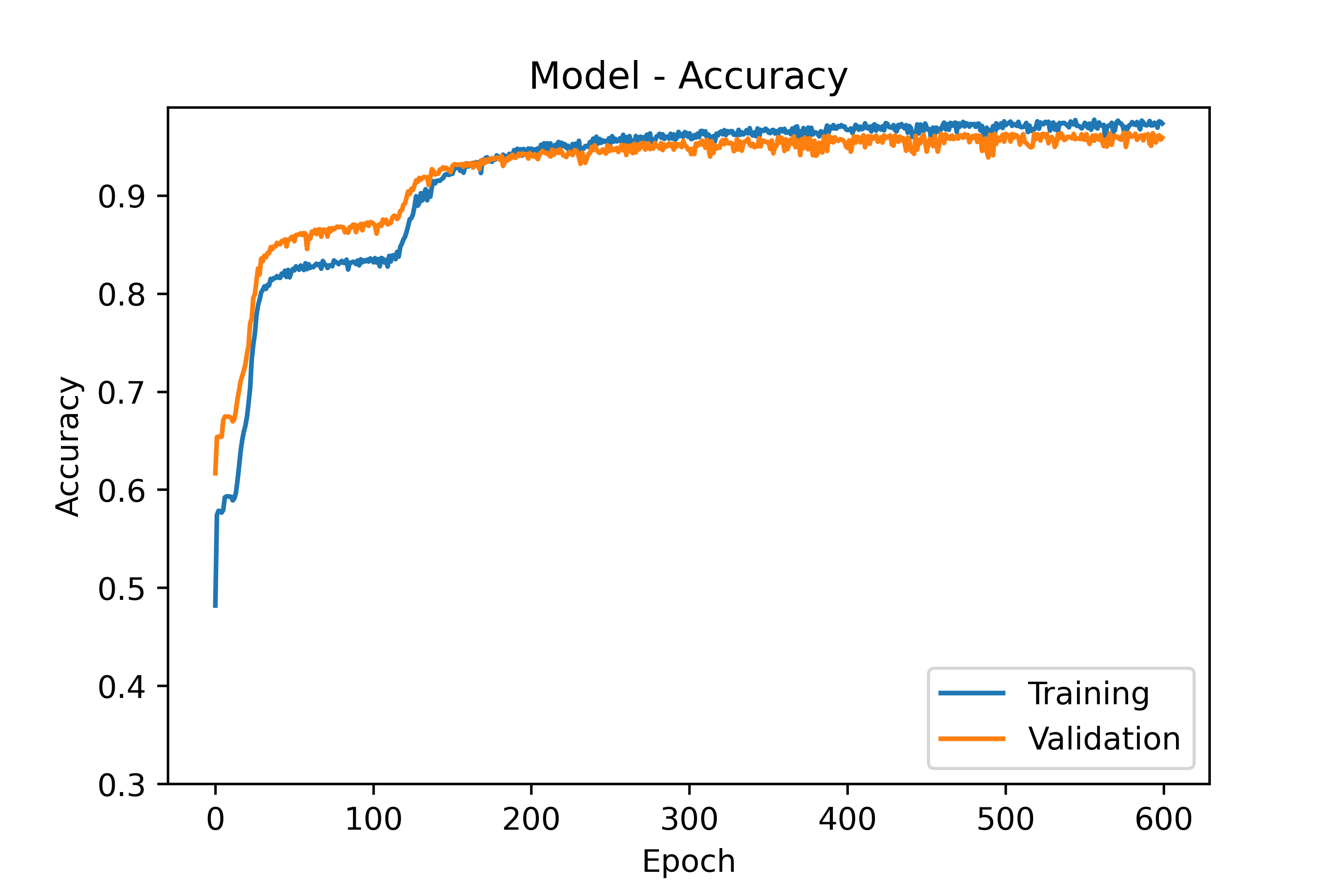}}
    \subfloat[]{\includegraphics[scale=.45]{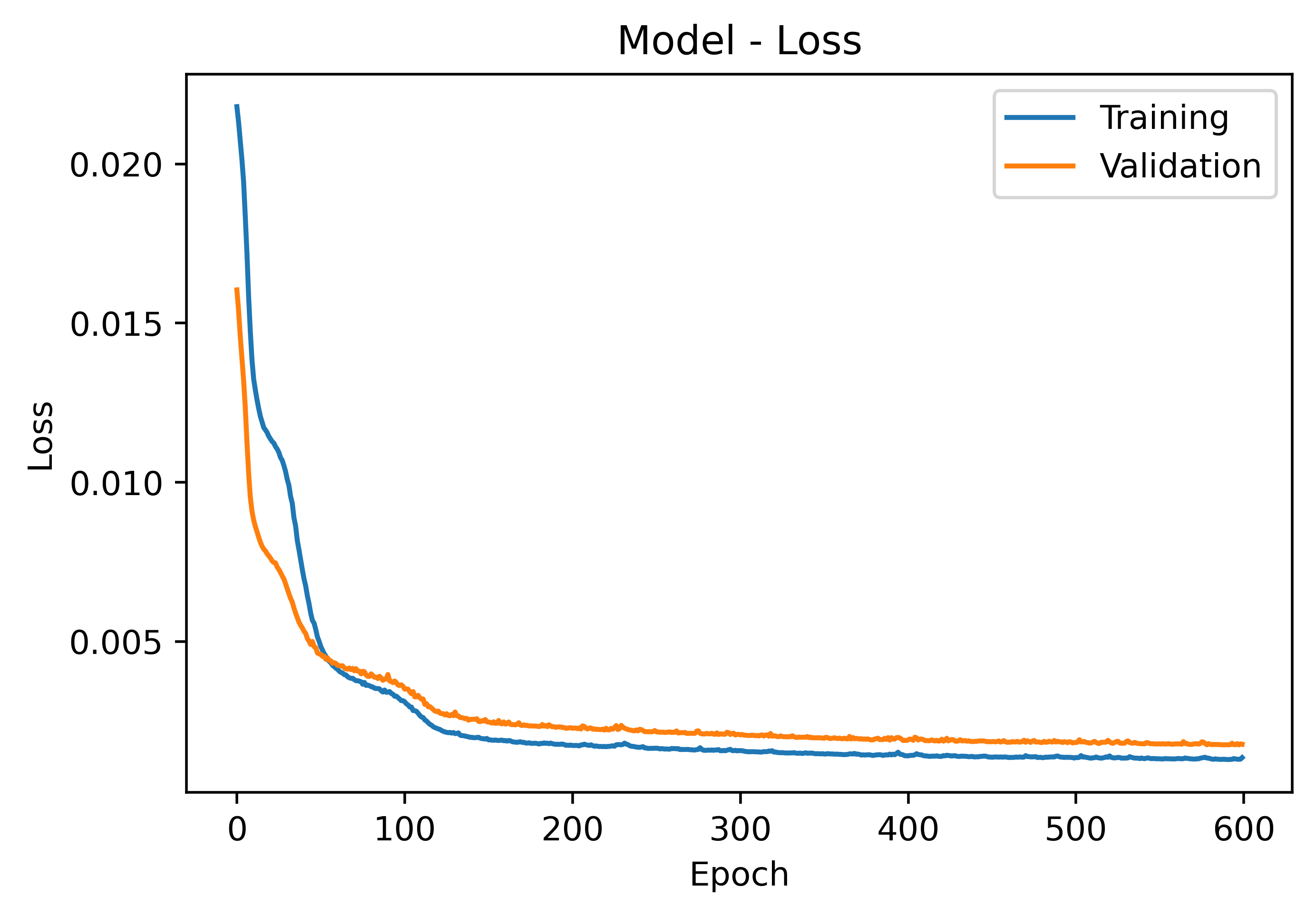}}
    \caption{{Model Training Parameters. (a) Autoencoder Training Accuracy Graph. (b) Autoencoder Training Loss Graph }}
    \label{Model graph}
\end{figure}

Autoencoders are artificial neural networks that can be used for image compression using unsupervised learning. If properly trained, they can be used to compress unseen images with an impressive compression ratio. Autoencoders are widely used for dimensionality reduction in generative models of data and contain three components; the encoder, latent vector and decoder. In the encoder, the size of the input is reduced by decreasing its dimension. The middle layer known as a latent vector has a few neurons as compared to the input and output layers. This layer holds the input information but in reduced dimensions named as the latent vector or shortcode. The output is reconstructed through the decoder using the reduced representation from the encoder output. Convolutional Autoencoders (CAE) effectively learn features of the input image in an unsupervised manner by employing a convolutional neural network (CNN). Convolution autoencoders can reconstruct the input image with minimum error. Once the autoencoder model is trained, it can generate a compressed representation of an unseen input image.\\
\begin{figure}[!t]
\captionsetup{justification=centering}
\centering\includegraphics[scale=.60]{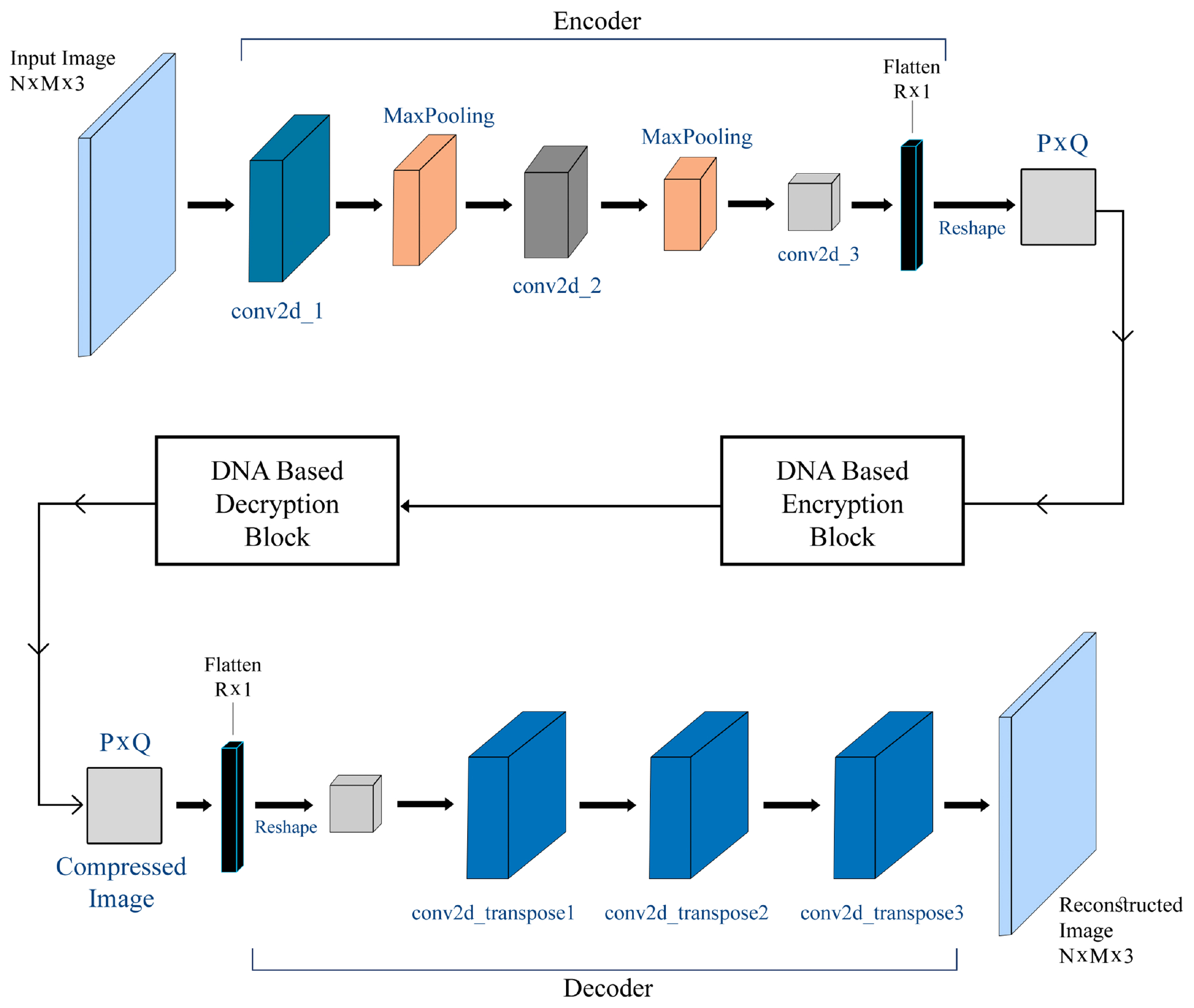}
\caption{Detailed Architecture of the Proposed Autoencoder along with DNA-based Encryption and Decryption Modules.}
\label{proposedschemedetail}
\end{figure}
\\
\newpage
\noindent\textbf{Autoencoder Architecture}
\\The proposed autoencoder architecture along with the encryption and decryption modules is shown in Fig.~\ref{proposedschemedetail}. The first part is the encoder in which an input RGB image of size $N \times M \times 3$ (512$\times$786$\times$3) is given as input to the encoder and a gray-scale image having dimensions $P \times Q$ (384$\times$512) is obtained at the output of the encoder. This compressed two-dimensional image can then be used by the encryption module instead of using the original colour image. This greatly helps to improve encryption and decryption speed as it is done in a lower dimension space. The encrypted image is obtained by employing four key-dependent different chaotic maps to randomize the process of permutation and diffusion. The same key is used at the decryption end to successfully recover the gray-scale image. The decrypted image is then given to the decoder module of the autoencoder model. If decryption is correctly done, the reconstructed image from the decoder is perceptually similar to the original colour image with the same dimension. Table \ref{summary} shows the architectural details of the proposed autoencoder model. {The experimental setup for implementing the autoencoder consisted of Google Colab with 12 GB NVIDIA Tesla K80 GPU and 12 GB ram. Moreover, Keras-based Application Programming Interface (API); v2.10 is used with TensorFlow v2.7 to create deep learning architecture, and OpenCv2-v4.5.5 is used for image dimension correction and normalization in this paper.}  \\
\begin{table}[!ht]
\begin{center}
\caption{\label{summary}Convolution Autoencoder Architecture Details.}
\scalebox{0.95}{
  \begin{tabular*}{\textwidth}{c @{\extracolsep{\fill}}ccccccl}
    \toprule
    \multicolumn{2}{c} {\textbf{Neural Network Layer}} & \textbf{The Feature Maps} &   \textbf{Size} & \textbf{Kernel} & \textbf{Stride} & \textbf{Activation Functions} \\
    \midrule
    Input Layer & Input Image & - & 512$\times$768$\times$3 & - & - & - \\
1	&Conv2d\_1	&128	& 512$\times$768$\times$128	&3x3 &1	&ReLU \\
2	& Max Pooling 2D	&128  &256$\times$384$\times$256	&2x2x2	&1	&- \\
3	&Conv2d\_2	&64	&256$\times$384$\times$64&	3x3	&1	&ReLU\\
4	&Max Pooling 2D	&64	&128$\times$192$\times$128	&2x2x2	&1	&-\\
5	&Conv2d\_3	&32	&64$\times$96$\times$32	&3x3	&2	&ReLU\\
6   &Flatten &-	&196608$\times$1	&-	&-	&-\\
7   &Encoder-Out &-	&384$\times$512	&-	&-	&-\\
8   &Decoder-In &-	&384$\times$512	&-	&-	&-\\
9  &Flatten &-	&196608$\times$1	&-	&-	&-\\
10	&Reshape &-	&64$\times$96$\times$32 &3x3 &-	&-\\
11	&Conv2d\_transpose1 &32	&128$\times$192$\times$32 &3x3 &2	&ReLU\\
12	&Conv2d\_transpose2	&64	&256$\times$384$\times$64 &3x3 &2	&ReLU\\
13	&Conv2d\_transpose3	&3	&512$\times$768$\times$3& 3x3& 2	&Sigmoid\\
Output &Image &- &512$\times$768$\times$3 &- &-	&-\\
  \bottomrule
\end{tabular*}}
\end{center}
\end{table}
\medskip
\newline\indent Coloured input images of size 512$\times$786$\times$3 are fed to the auto-encoder model. Convolution filters of sizes 128, 64, and 32 are used for the encoder part to extract spatial information. The dimension of the input layer is reduced by using Max Pooling. Except for the last layer of the encoder part, all other convolutional layer strides are of size 1. Padding of the same size is used for all the layers. The output after conv2d\_3 is $64 \times 92 \times 32$ which contains 32 feature maps but to get a latent vector, this layer has been flattened to produce a vector of $R \times 1$ dimension. For the specific colour image size considered in this paper, R is a latent vector of size 196608. This vector is then reshaped to obtain a 2-dimensional gray-scale image $P \times Q$ (384$\times$512). The compression ratio for this work is 6:1. The gray-scale image is encrypted using the proposed encryption algorithm. Once the gray-scale image is decrypted, it is then given to the decoder part of the autoencoder which uses transpose convolutional filters to regenerate the original input colour image with negligible loss of information. The proposed CAE model is very lightweight. In terms of trainable parameters, the encoder has 95,840 and the decoder has 29,475 parameters. Due to low training parameters, the compression time is very low for the proposed convolutional autoencoder technique.\\
\newline\noindent\textbf{{Autoencoder Training}}
\\The proposed convolutional autoencoder-based deep learning model was trained for different epochs, hyper-parameters and optimizers using the Kodak dataset which has 24 uncompressed similar dimension images. The dataset was divided into a 60\% training set and a 40\% validation set. The model converged near 550 epochs at the maximum training accuracy of 97\% and a loss of 0.0011, whereas the validation accuracy and loss were 95\% and 0.0014, respectively as shown in Fig.~\ref{Model graph}. To avoid overfitting, kernel and bias regularization techniques have been used in all convolution layers for the autoencoder model. L2 regularization with a specific value of $1e^{-6}$ is used to reduce the size of weights, as well as the biases during model training. This helps to avoid overfitting and also removes the fluctuations in model performance metrics. L2 regularization forces small weights to be removed that are very close to zero. This technique improves the model performance.\\

\begin{figure}[!ht]
  \centering
    \captionsetup{justification=centering}
    \subfloat[]{\includegraphics[scale=.25]{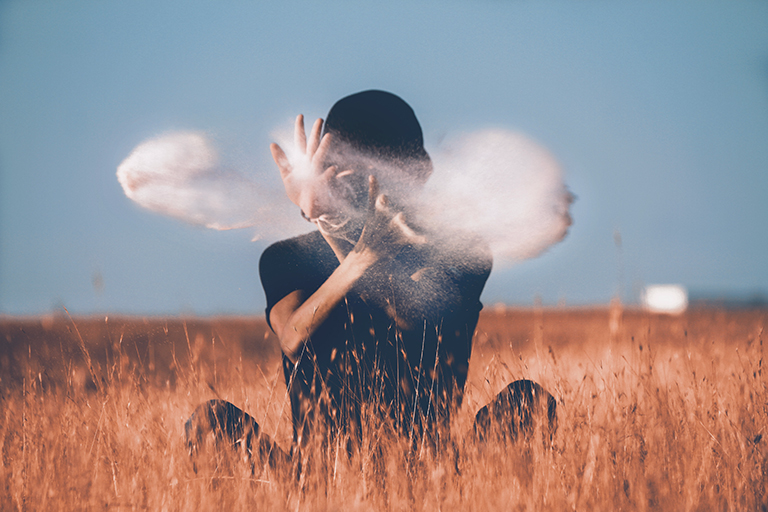} }
    \subfloat[]{\includegraphics[scale=.25]{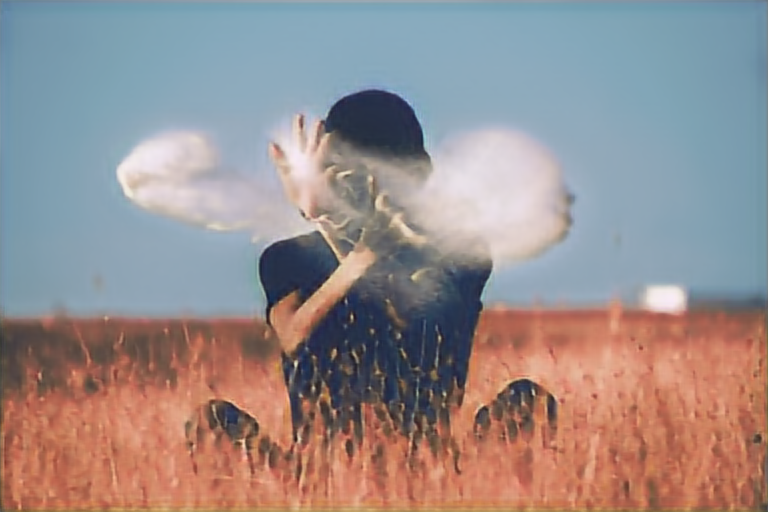}}
    \caption{Field Image. (a) Original Field Image. (b) Reconstructed Field Image}
    \label{field}
\end{figure}

\begin{figure}[!ht]
  \centering
    \captionsetup{justification=centering}
    \subfloat[]{\includegraphics[scale=.25]{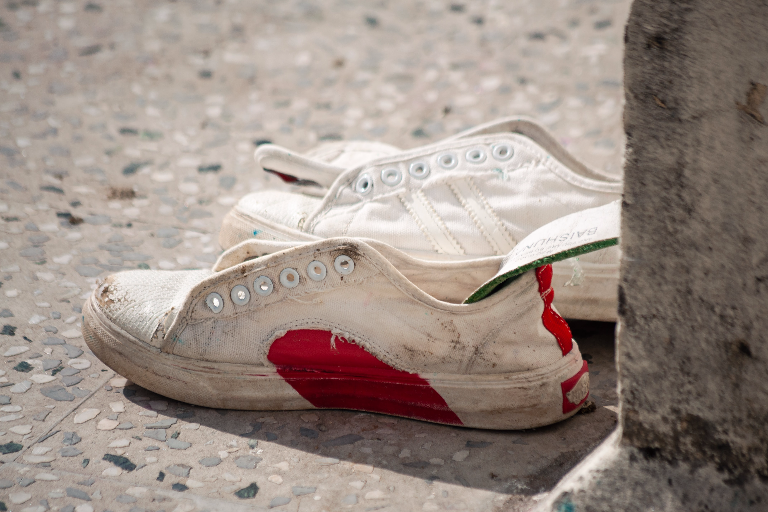}}\hspace{0.3em}
    \subfloat[]{\includegraphics[scale=.25]{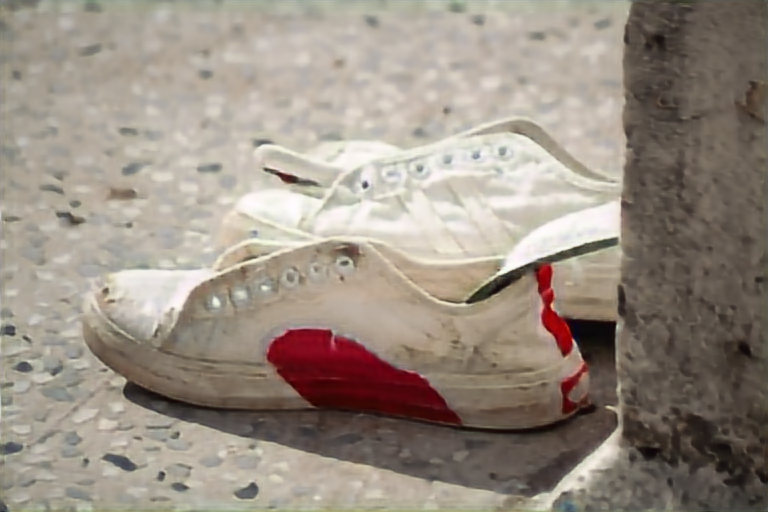}}
    \caption{Shoe Image. (a) Original Shoe Image. (b) Reconstructed Shoe Image. }
    \label{shoes}
\end{figure}

\noindent\textbf{{Autoencoder Tuning}}
\\Three types of optimizers were tested during the training of the proposed autoencoder model with different hyperparameter values. The Adam optimizer with an adaptive learning rate was initially selected, but it took very long to improve the training accuracy, therefore, a learning rate of $1e^{-3}$ was used which improved the training speed, but the model accuracy did not improve beyond 88\%. The RMSprop optimizer was then used as a replacement for the Adam optimizer with a learning rate of $1e^{-4}$. The model accuracy improved to 90\% but still as compared to other research, this accuracy was not enough. The Adamax optimizer with mean-squared-error loss function, $1e^{-4}$ learning rate and 600 epochs, achieved 95\% validation accuracy. Data related to model training is presented in Table~\ref{opti}.\\
\begin{table}[!h]
\begin{center}
\caption{\label{opti} Experimenting Hyper-parameters and Optimizers.}
\scalebox{0.98}{
  \begin{tabular}{cccc}
    \toprule
    \textbf{Hyper-Parameters}  \\
    \toprule
    Optimizer & Learning Rate & Epochs & Validation Accuracy \\
    \midrule
    Adam & 0.001 & 500 & 88\% \\
    RMSprop & 0.0001 & 300 & 90\% \\
    Adamax & 0.0001 & 600 & 95\% \\
  \bottomrule
\end{tabular}}
\end{center}
\end{table}
\medskip
\newline\noindent\textbf{{Autoencoder Performance}}
\\Many researchers have designed deep convolutional neural network-based image compression techniques using different datasets such ImageNet's large database or Kodak's small dataset and various input image sizes. The authors in \cite{cheng2018deep} claim that their compression technique is close to JPEG2000 but this technique is used for small grayscale images of size 128$\times$128. If the input image size is greater than 128$\times$128, then the large size image will split into non-overlapping small images and each of these samples will be compressed separately. The technique proposed in \cite{cheng2018deep} has produced better results than \cite{balle2016end} using the same database but the model was tested on the Kodak database yielding 13.7\% BD-rate saving average on all 24 images. The PSNR of the reconstructed images for the same Kodak database is between 29-33 dB.\\
\\
\noindent\textbf{Quality of the Reconstruct Images using the Proposed Autoencoder Module}
\\Using the proposed autoencoder, a few sample images are shown from the Kodak dataset. Table \ref{PSNR} shows the  PSNR values of each image. Specifically, the PSNR value of the Building image is 26 dB, whereas the Field and Shoe images illustrated in Fig.~\ref{field} Fig.~\ref{shoes} have PSNR values of 30 and 31 dB, respectively.
\begin{figure}[!ht]
  \centering
    \captionsetup{justification=centering}
    \subfloat[]{\includegraphics[scale=.25]{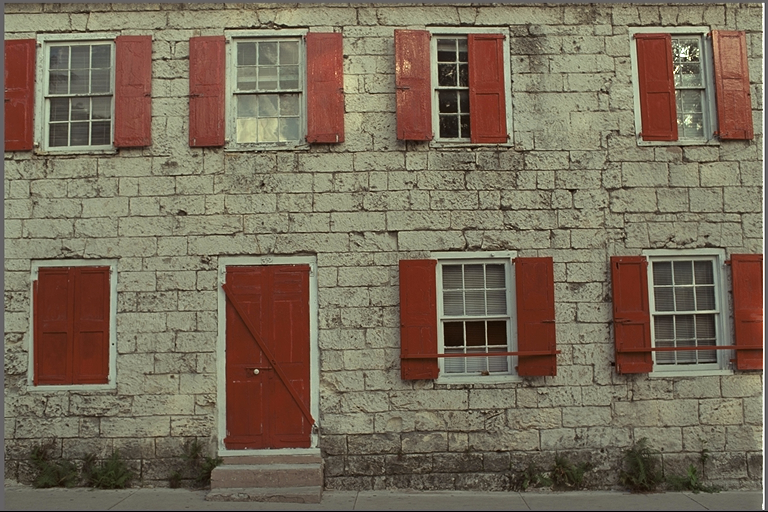} }
    \subfloat[]{\includegraphics[scale=.25]{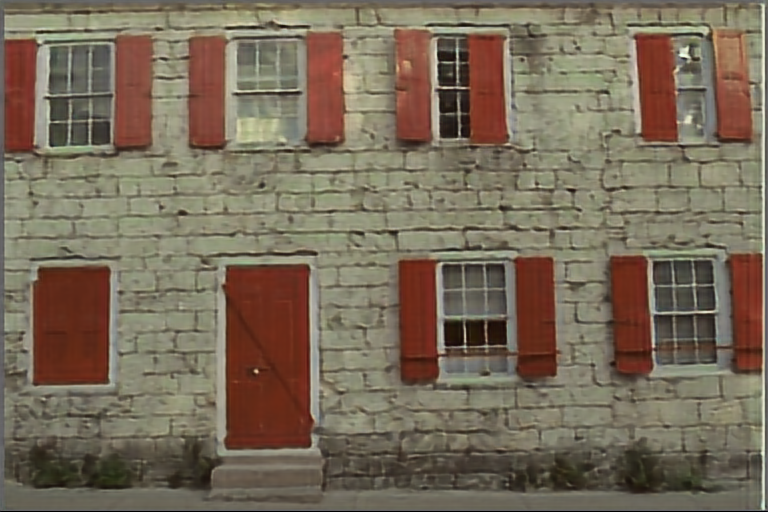}}
    \caption{Building Image. (a) Original Building Image. (b) Reconstructed Building Image.}
    \label{build}
\end{figure}
\vspace{10mm}
\begin{table}[!ht] 
\begin{center}
\caption{\label{PSNR} PSNR Table for Sample Images.} 
\scalebox{0.99}{
\begin{tabular}{ cc }
\toprule
Images & PSNR(dB)  \\
\midrule
Building Image & 26  \\
Field Image & 30 \\
Shoes Image & 31\\
\bottomrule
\end{tabular}}
\end{center}
\end{table}
\subsection{The Proposed DNA based Colour Image Encryption Scheme}
Several image encryption schemes employing chaos theory can be found in the literature. Due to sensitivity to initial conditions, chaotic maps produce a secure ciphertext image. But many of these schemes have been cracked and cryptanalysis of such schemes is reported in the literature. {Therefore, the encryption scheme proposed in this paper combines chaos and DNA for a high level of security. The proposed encryption scheme has been implemented in MATLAB R2020b on an Intel(R) Core(TM) i7 processor with 16GB of ram. } 
%%%%%%%%%%%%%%%%%%%%%%%%%%%%%%%%%5
\iffalse
%%%%%%%%%%%%%%%%%%%%%%%%%%%%
%\begin{equation} \label{discrete_LMAP}
%x_{n+1}= rx(1-x_{n}),   \mbox{    }   0\le %x_{n} \le 1.
%\end{equation}
%%%%%%%%%%%%%%%%%%%%%%%%%%%%
\begin{figure}[!ht]
  \centering
  \subfloat[]{\includegraphics[width=.40\columnwidth]{r1.png} }
  \subfloat[]{\includegraphics[width=.40\columnwidth]{r15.png}  }\\
  \subfloat[]{\includegraphics[width=.40\columnwidth]{r20.png} }
  \subfloat[]{\includegraphics[width=.40\columnwidth]{r25.png}  }\\
   \subfloat[]{\includegraphics[width=.40\columnwidth]{r30.png} }
  \subfloat[]{\includegraphics[width=.40\columnwidth]{r35.png}  }\\
   \subfloat[]{\includegraphics[width=.40\columnwidth]{r36.png} }
  \subfloat[]{\includegraphics[width=.40\columnwidth]{r37.png}  }\\
  \caption{Plot of Logistic Map Against a Number of Iterations for Different Values of $r$ ($x_{0} = 0.1$).}
  \label{fig:logistic_orbit}
\end{figure}
\fi
%%%%%%%%%%%%%%%%%%%%%%%%%%%%5

This section presents the workflow of the proposed colour image encryption scheme. The overall process is illustrated in Fig.~ \ref{encryp} and various steps are outlined as follows:  
\begin{figure}[!ht]
\captionsetup{justification=centering}
\centering\includegraphics[scale=.60]{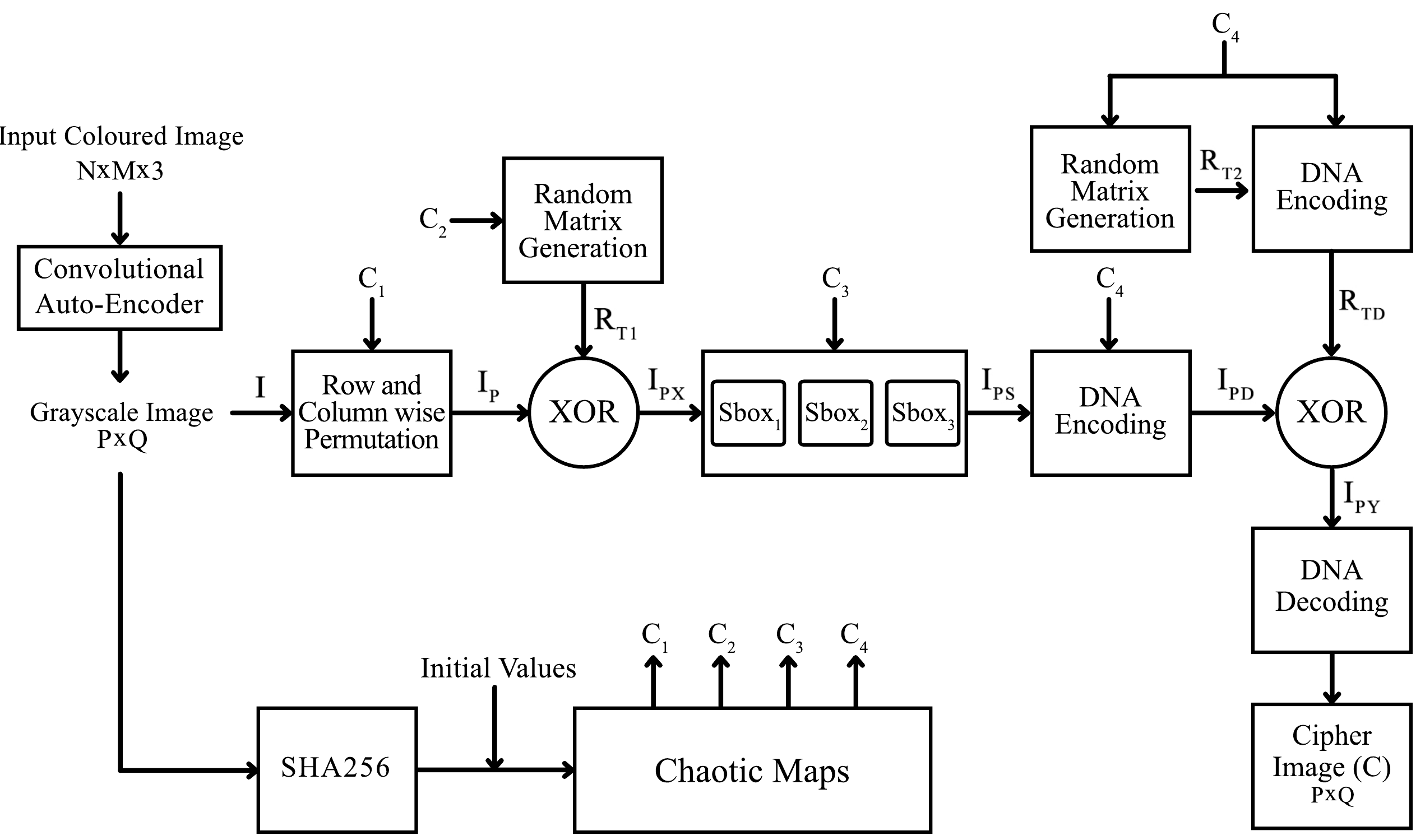}
\caption{An All-inclusive Flow of the Proposed Encryption Scheme.}
\label{encryp}
\emph{}
\end{figure}

\begin{figure}[!ht]
  \centering     
  \subfloat[ ]{\includegraphics[width=.3\textwidth]{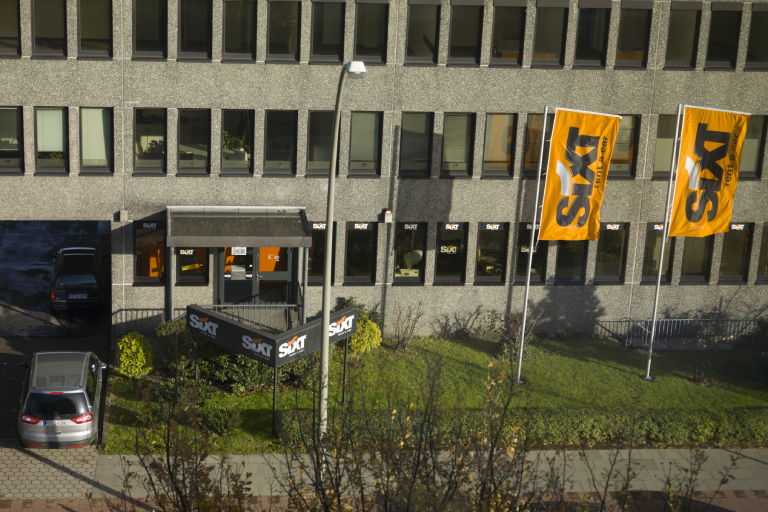} } \\ \hspace{4mm}             
  \subfloat[ ]{\includegraphics[width=.3\textwidth]{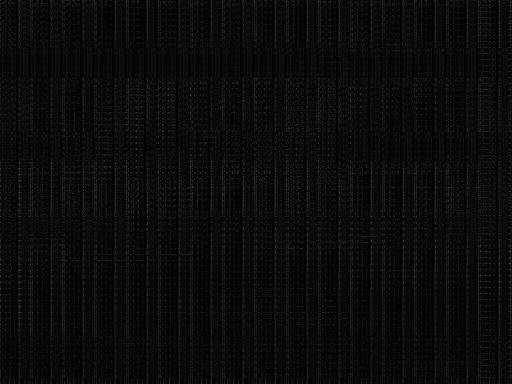} } \hspace*{0.5cm} 
  \subfloat[]{\includegraphics[width=0.3\textwidth]{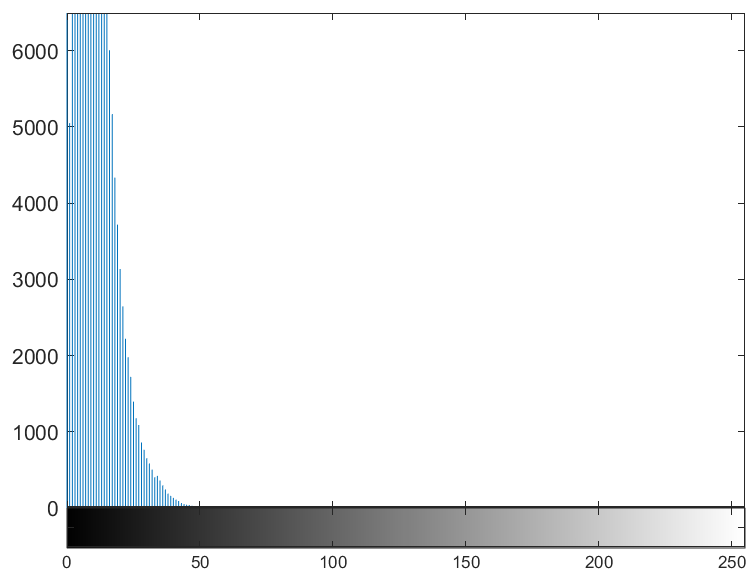} } \\ \hspace*{0.5cm}  
  \subfloat[ ]{\includegraphics[width=0.3\textwidth]{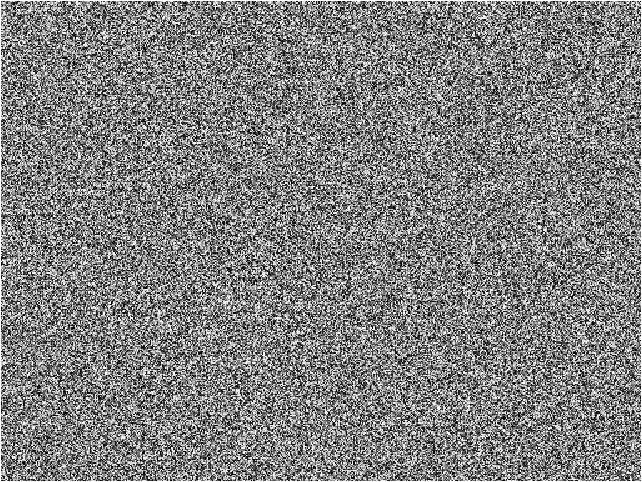} }
  \hspace*{0.4cm} 
  \subfloat[]{\includegraphics[width=0.3\textwidth]{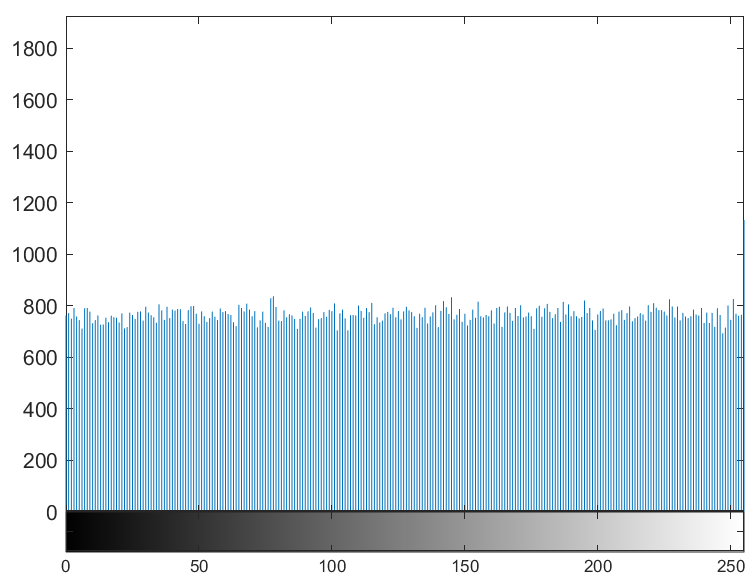} }
  
  \caption{Original Image, Histogram and Cipher Image: The Building Image. (a) Original building image. (b) The Encoded variant. (c) The histogram of the encoded variant. (d) The Encrypted image. (e) The histogram of the encrypted image. }
  \label{Enc1}
\end{figure}
%commented below
\iffalse
\begin{figure}[!ht]
  \centering     
  \subfloat[Original image.  ]{\includegraphics[width=.3\textwidth]{image2-org.png} } \\ \hspace{4mm}              
  \subfloat[Encoded image. ]{\includegraphics[width=.3\textwidth]{image2-en.jpeg}  } \hspace*{0.5cm} 
  \subfloat[Histogram of the encoded image. ]{\includegraphics[width=0.3\textwidth]{hist2.png} } \\ \hspace*{0.5cm}  
  \subfloat[Encrypted image. ]{\includegraphics[width=0.3\textwidth]{Encr2.png} }
  \hspace*{0.5cm} 
  \subfloat[Histogram of the encrypted image. ]{\includegraphics[width=0.3\textwidth]{hist2-e.png} }
  
  \caption{Original image, Histogram and Cipher Image: The Shoe Image}
  \label{Enc2}
\end{figure}

\begin{figure}[!ht]
  \centering     
  \subfloat[Original image.  ]{\includegraphics[width=.3\textwidth]{image3-org.png} } \\              
  \subfloat[Encoded image. ]{\includegraphics[width=.3\textwidth]{image3-en.jpeg}  } \hspace*{0.5cm} 
  \subfloat[Histogram of encoded image. ]{\includegraphics[width=0.3\textwidth]{hist3.png} } \\ \hspace*{0.5cm}  
  \subfloat[Encrypted image. ]{\includegraphics[width=0.3\textwidth]{Encr3.png} }
  \hspace*{0.5cm} 
  \subfloat[Histogram of encrypted image. ]{\includegraphics[width=0.3\textwidth]{hist3-e.png} }
  
  \caption{Original image, Histogram and Ciphertext Image: Image 3. }
  \label{Enc3}
\end{figure}
\fi
\begin{figure}[!ht]
  \centering     
  \subfloat[ ]{\includegraphics[width=.3\textwidth]{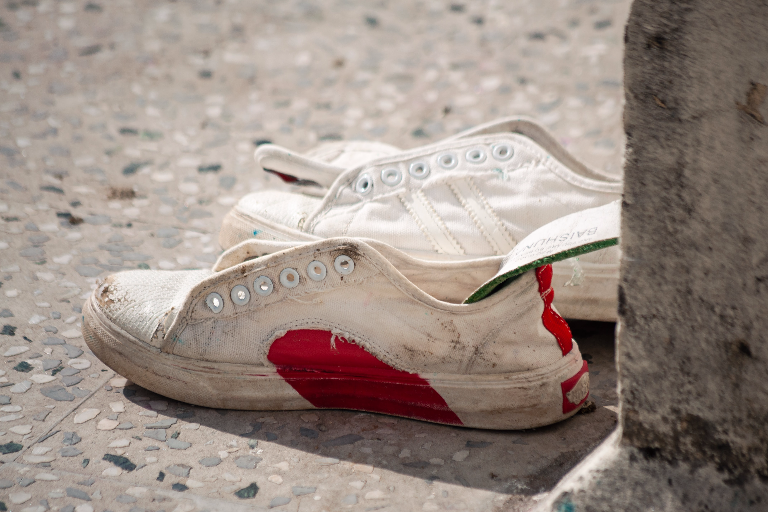} } \\              
  \subfloat[ ]{\includegraphics[width=.3\textwidth]{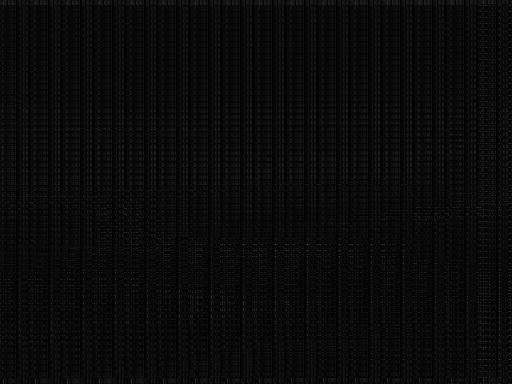}  } \hspace*{0.5cm} 
  \subfloat[ ]{\includegraphics[width=0.3\textwidth]{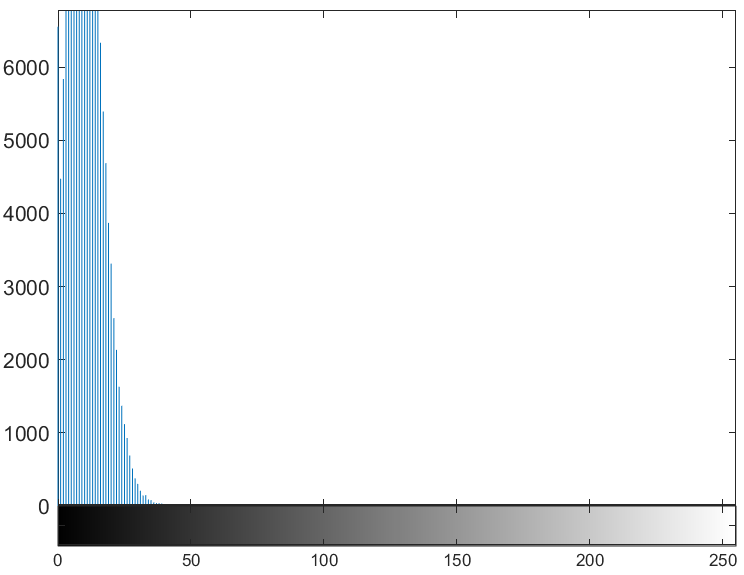} } \\ \hspace*{0.5cm}  
  \noindent\subfloat[ ]{\includegraphics[width=0.3\textwidth]{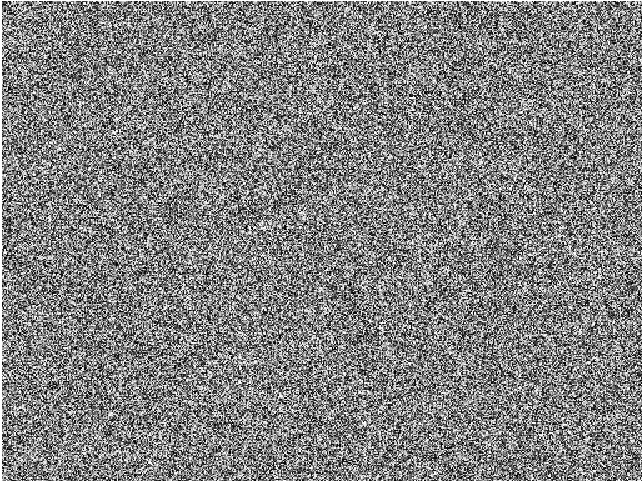} }
  \hspace*{0.5cm} 
  \subfloat[ ]{\includegraphics[width=0.3\textwidth]{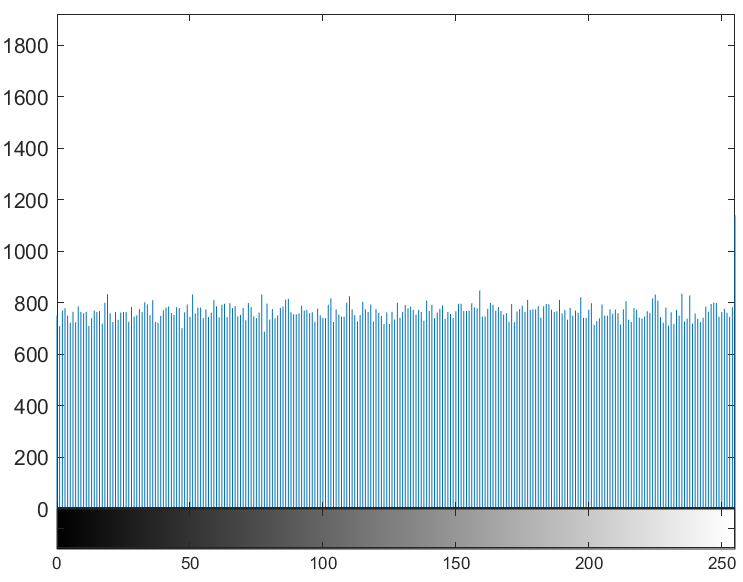} }
  
  \caption{Original Image, Histogram and Ciphertext Image: Shoe Image. (a) Original Shoe image. (b) The encoded variant. (c) Histogram of the encoded variant. (d) The encrypted image. (e) The histogram of the encrypted image. }
  \label{Enc4}
\end{figure}
%commented below
\iffalse
\begin{figure}[!ht]
  \centering     
  \subfloat [ ]{\includegraphics[width=.3\textwidth]{image5-org.JPG} } \\              
  \subfloat[ ]{\includegraphics[width=.3\textwidth]{image5-en.jpeg}  } \hspace*{0.5cm} 
  \subfloat[ ]{\includegraphics[width=0.3\textwidth]{hist5.png} } \\ \hspace*{0.5cm}  
  \subfloat[ ]{\includegraphics[width=0.3\textwidth]{Encr5.png} }
  \hspace*{0.5cm} 
  \subfloat[ ]{\includegraphics[width=0.3\textwidth]{hist1-e.png} }
  
  \caption{Original Image, Histogram and Ciphertext Image: Image 5. (a) Original image. (b) Encoded image. (c) Histogram of the encoded image. (d) Encrypted image. (e) Histogram of the encrypted image.  }
  \label{Enc5}
\end{figure}
\fi

\begin{figure}[!ht]
  \centering     
  \subfloat[  ]{\includegraphics[width=.3\textwidth]{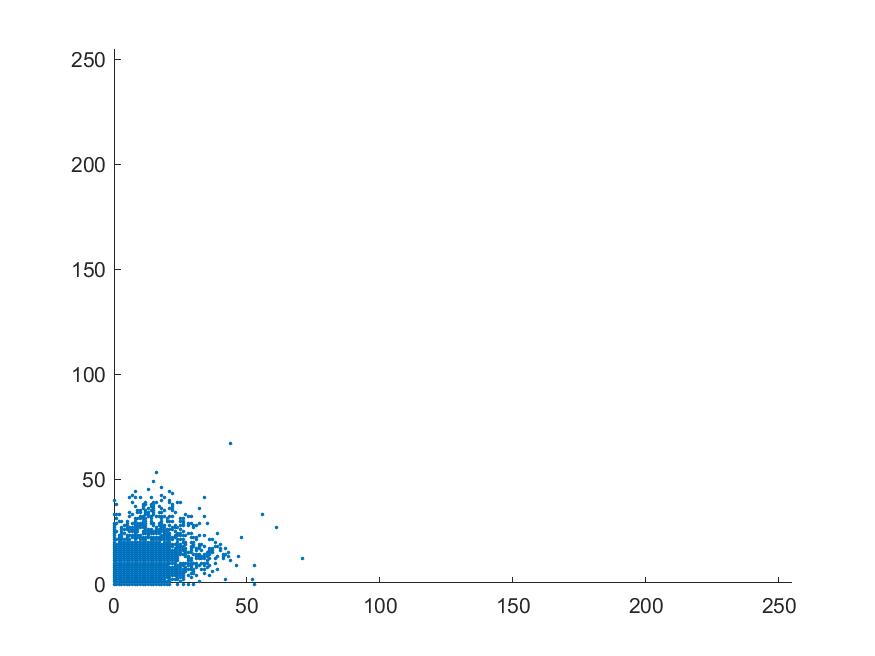} }   
  \subfloat[ ]{\includegraphics[width=.3\textwidth]{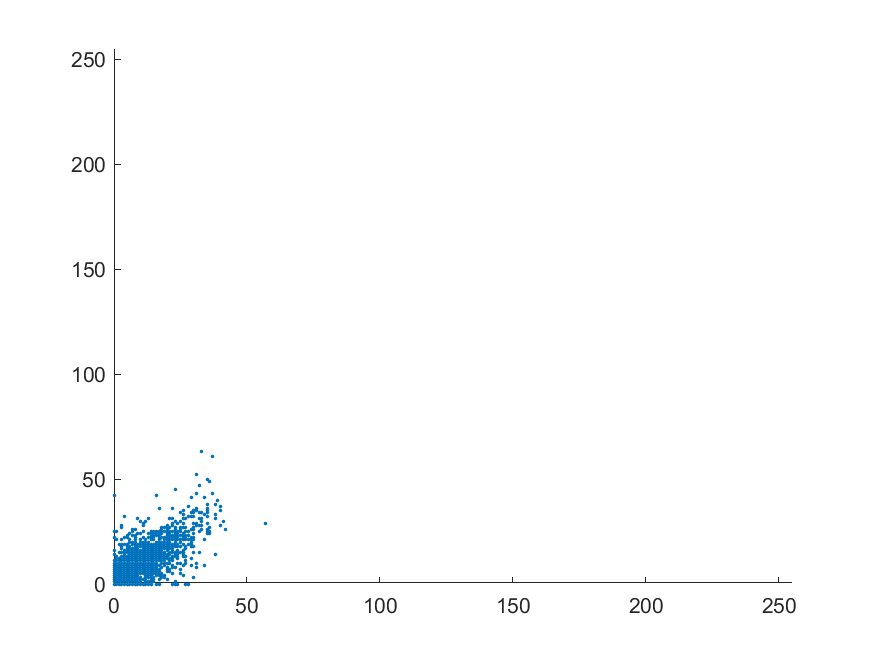}  } 
  \subfloat[ ]{\includegraphics[width=.3\textwidth]{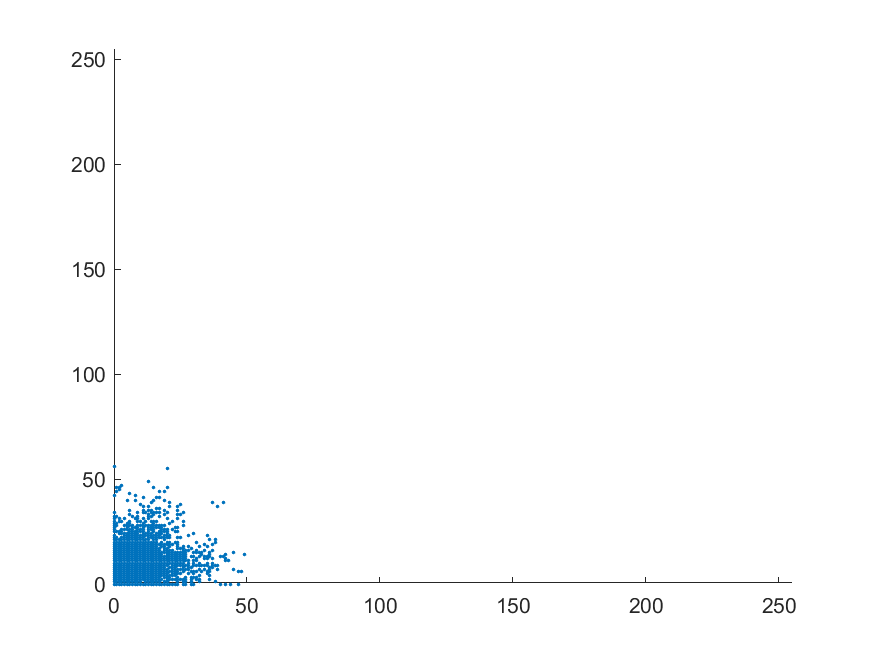}  } \\ 
  \subfloat[ ]{\includegraphics[width=0.3\textwidth]{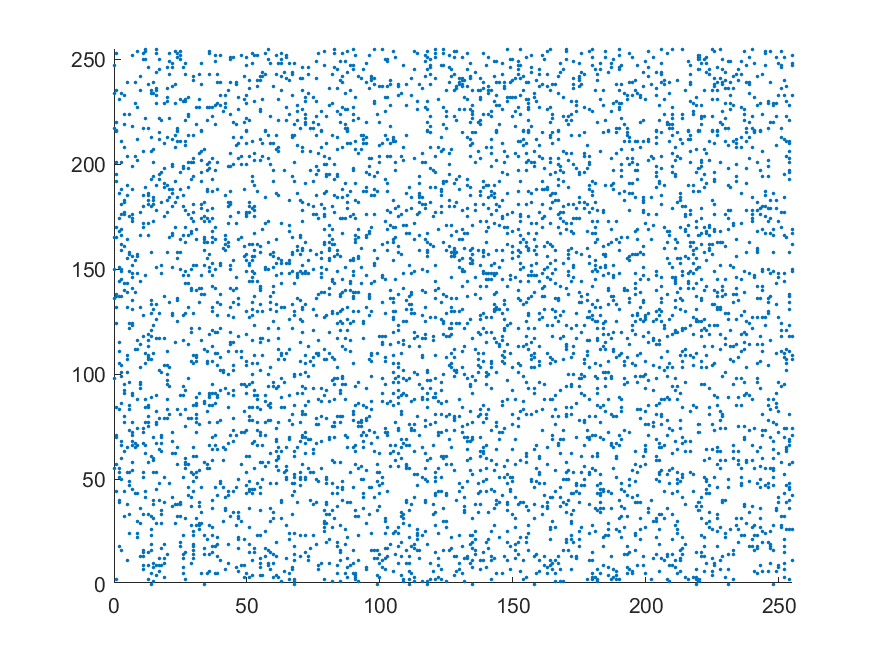} }   
  \subfloat[ ]{\includegraphics[width=0.3\textwidth]{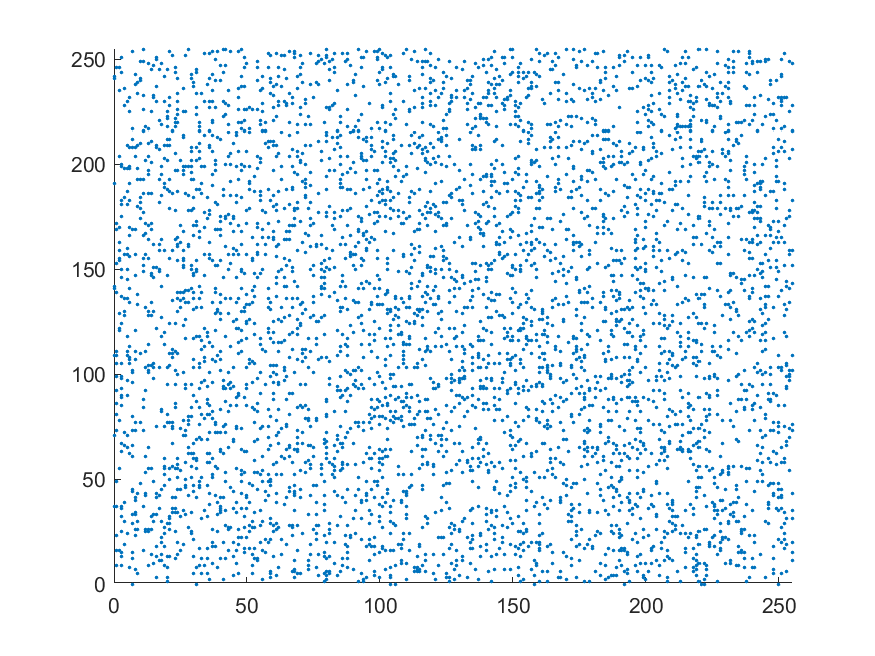} }
  \subfloat[ ]{\includegraphics[width=0.3\textwidth]{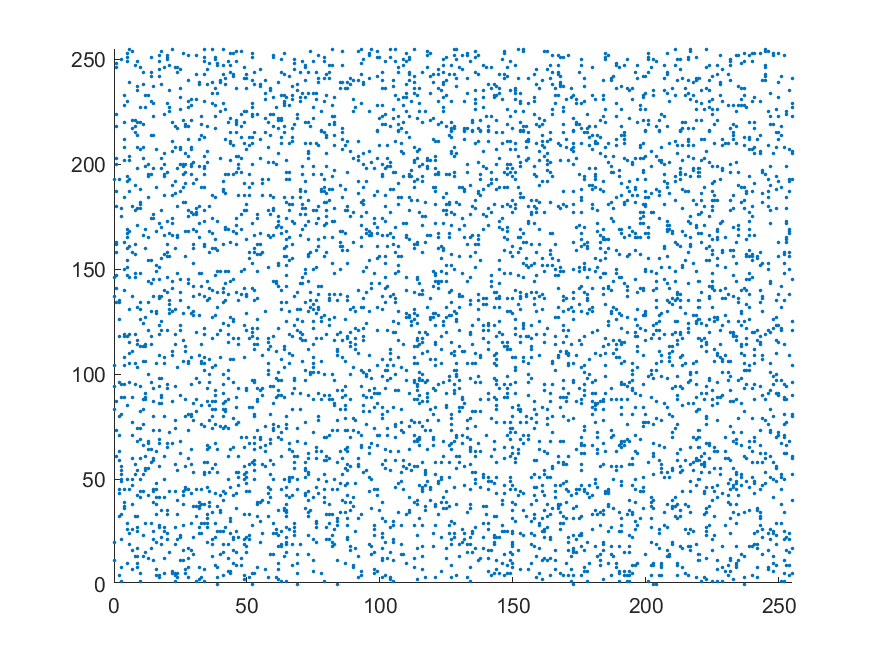} } \\
  %%%%%%%%%%%%%%%%%%%%%%%% image 2
  \subfloat[ ]{\includegraphics[width=.3\textwidth]{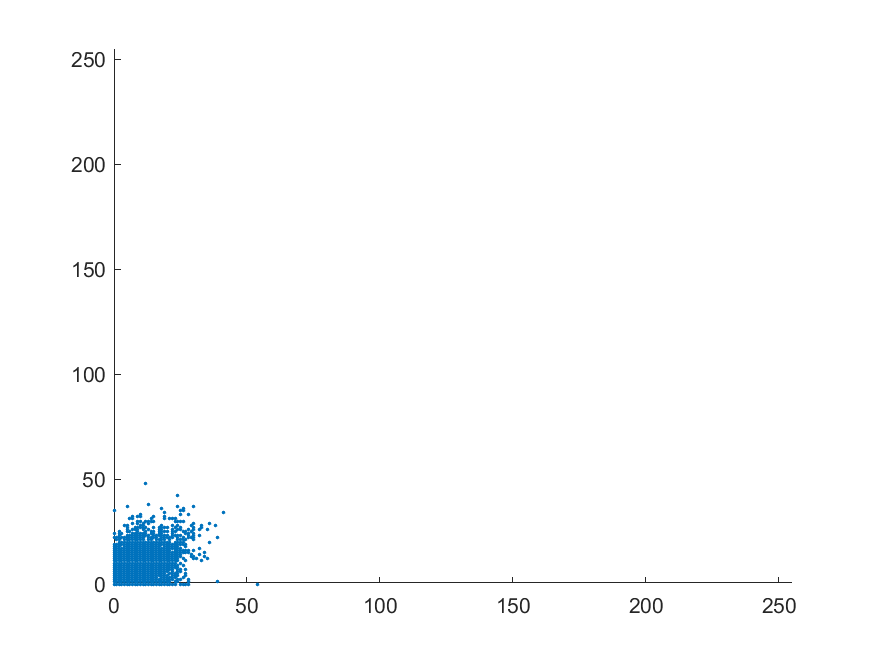} }      
  \subfloat[ ]{\includegraphics[width=.3\textwidth]{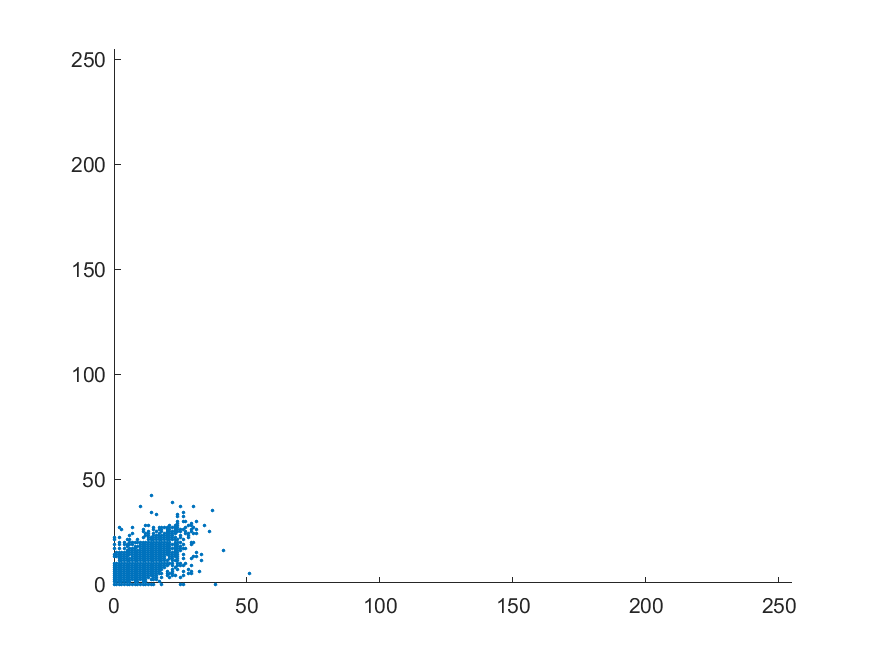}  } 
  \subfloat[ ]{\includegraphics[width=.3\textwidth]{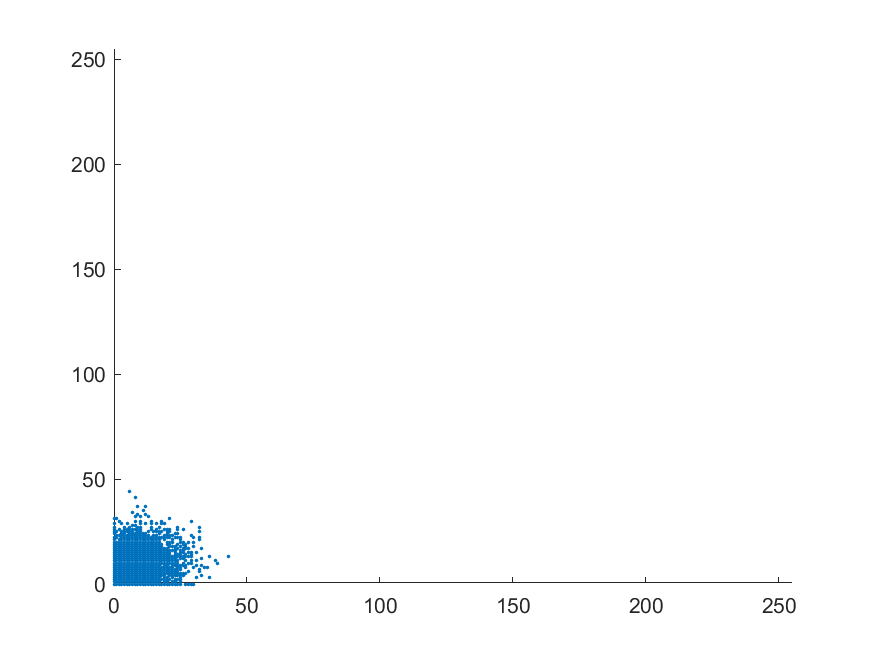}  } \\ 
  \subfloat[ ]{\includegraphics[width=0.3\textwidth]{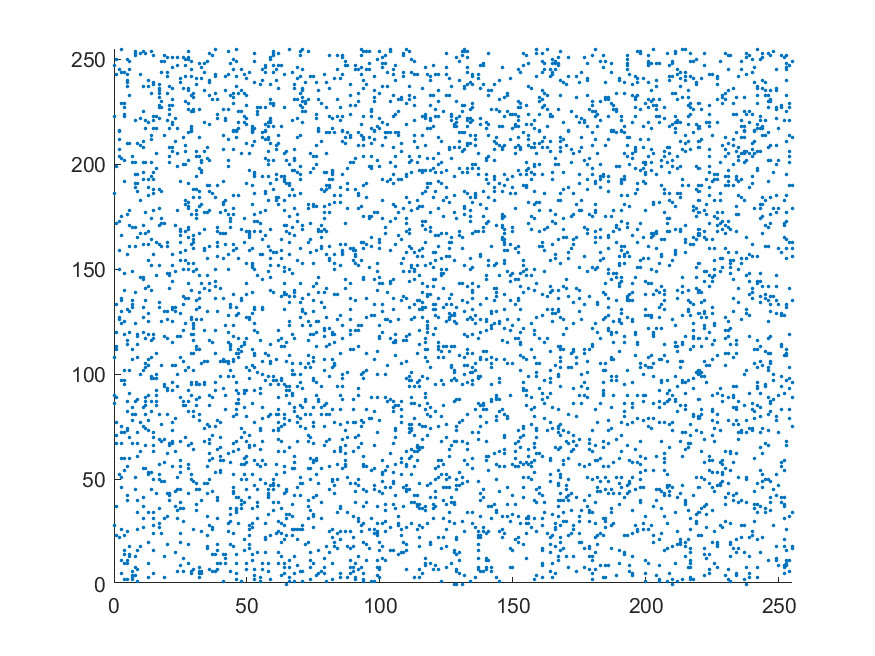} }  
  \subfloat[ ]{\includegraphics[width=0.3\textwidth]{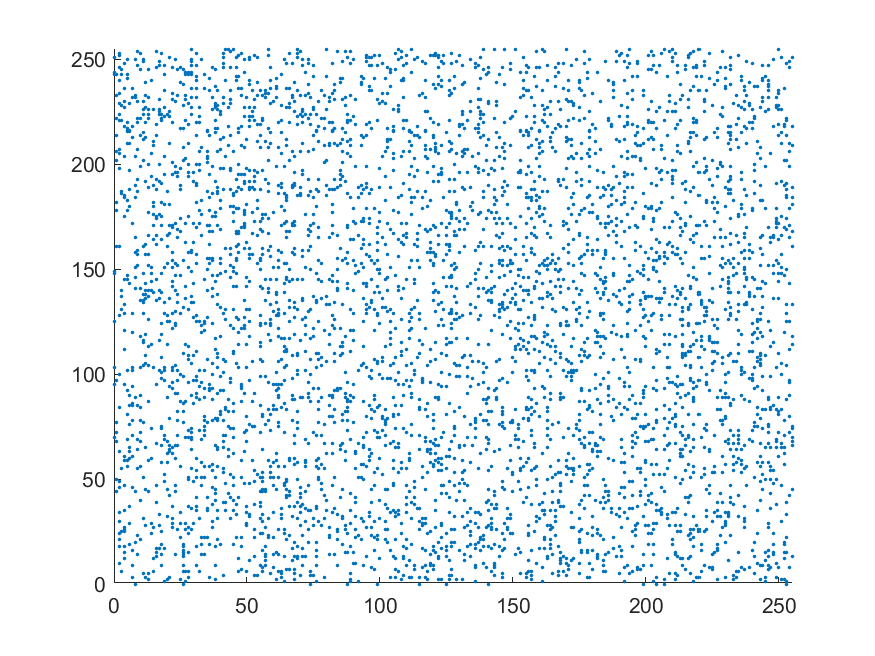} }
  \subfloat[ ]{\includegraphics[width=0.3\textwidth]{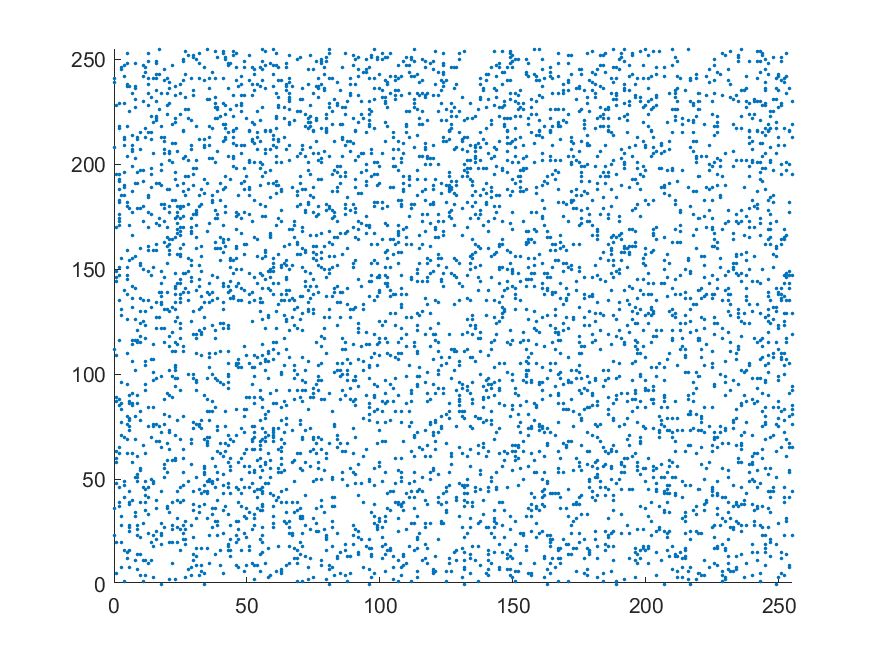} }
  
  \caption{Correlation Plots of the Original and Cipertext Images; (a-c) Original Image 1, (d-f) Encrypted Image 1, (g-i) Original Image 2, (j-l) Encrypted image 2.}
  \label{corr-image}
\end{figure}

%%%%%%%%%%%%%%%%%%%%%%%%%%%%%%%%%%%%%%%
\begin{table}[!ht]
\center
\setlength{\tabcolsep}{0.4em}
\renewcommand{\arraystretch}{1.8}
\caption{Security Analysis of the Encryption Results}
\label{tab:enc}       % Give a unique label
\scalebox{1.2}{
\begin{tabular}{lllll}
\hline
Security Metric & Plaintext-Image & Encrypted-Image & Plaintext-Image & Encrypted-Image  \\
\hline
Horizontal Correlation Coefficient &   0.9612 &  0.0021  &   0.9751 &  0.0026  \\
Vertical Correlation Coefficient &  0.9833 &   -0.0043 &   0.9712 &  0.0035 \\
Diagonal Correlation Coefficient & 0.9798  &  0.0033 &   0.9712 &  0.0027\\
E (Entropy)  &  7.34 & 7.99 &  7.39 &  7.99 \\
KS (Key Sensitivity)  & NA & 99.33\% &   NA &  99.89\%  \\
NPCR  & NA & 99.78 \%  &   NA &  99.85\%\\
UACI  & NA & 34.25  &   NA &  35.11\\
C (Contrast)  & 1.21 & 10.48 &   1.24 &  10.45 \\
H (Homogeneity)  & 0.71 & 0.37 &   0.72 &  0.36 \\
E (Energy)  & 0.11 & 0.02 &   0.12 &  0.03 \\
\hline
\end{tabular}}
\end{table}
\vspace{6mm}
\begin{enumerate}
    \item The input colour image having a dimension of N$\times$M$\times$3 is first reduced to a smaller dimension, P$\times$Q, thus forming a gray-scale image $I$ using the proposed convolutional autoencoder module.
    \item Hash of the image $I$ is then taken using SHA-256 to obtain a key $K$ that is sensitive to $I$. Initial conditions are generated using $K$ and the initial values $\mu_0$ for the TD-ERCS chaotic map, $X_0$ for the Intertwining chaotic map, $C_0$ for the NCA map, and $A_0$ for the Chirikov chaotic map. The chaotic maps, TD-ERCS, Intertwining, NCA and Chirikov are referred to as $C_1$,$C_2$, $C_3$ and $C_4$, respectively.
    
    \item Randomly permutes rows of the image $I$ using the TD-ERCS map $C_1$ to obtain an intermediate image $I_{PR}$.
    
    \item Randomly permute columns of the image $I_{PR}$ using the TD-ERCS map $C_1$ to obtain the permuted image $I_P$.
    
    \item A random matrix is generated using the Interwinning map $C_2$. Since the values of the random matrix are small, therefore, to scale these values between 0 to 255, multiply each entry of the random matrix by $10^{14}$ and take modulo 256 to get values in the range 0 to 255. Let this random matrix be referred to as $R_{T1}$.
    
    \item XOR $I_P$ and $R_{T1}$ to obtain $I_{PX}$. 
    
    \item For each pixel of $I_{PX}$, the NCA map $C_3$ is used to randomly select one of the three s-boxes. Each pixel value of $I_{PX}$ is then replaced with the corresponding value of the randomly selected S-box. After iterating through all the pixels of $I_{PX}$, the substituted image $I_{PS}$ is obtained.
    
    \item Convert the image $I_{PS}$ into a binary matrix and perform DNA encoding by randomly selecting one of the eight DNA rules shown in Table \ref{tab:dna code}. The Chirikov chaotic map $C_4$ is used for random selection of the rule from Table \ref{tab:dna code}. The DNA encoded image is referred to as $I_{PD}$.
    
    \item Generate a random matrix using the Chirikov chaotic map $C_4$. Since the values of the random matrix are small, therefore, to scale these values between 0 to 255, multiply each entry of the random matrix by $10^{14}$ and take modulo 256 to get values in the range 0 to 255. Let this random matrix be referred to as $R_{T2}$.

    \item Convert the random matrix $R_{T2}$ into a binary matrix and perform DNA encoding of each entry of $R_{T2}$ by randomly selecting one of the eight DNA rules shown in Table \ref{tab:dna code}. Furthermore, to select a random DNA code from Table \ref{tab:dna code}, the Chirikov chaotic map $C_4$ is used, which in turn generates the DNA encoded image $R_{TD}$.

    \item XOR $I_{PD}$ and $R_{TD}$ to obtain $I_{PY}$.
    
    \item The cipher image $C$ of dimension $P \times Q$ is obtained by performing DNA decoding of $I_{PY}$. 

\end{enumerate}
\section{Security Evaluation of the Proposed Scheme}\label{sec5}
Several security evaluation parameters, such as histogram of the cipher image, entropy, NPCR, UACI, key sensitivity, contrast, etc., are presented in this section to evaluate the efficacy of the proposed DNA encryption technique. 

\subsection{Histogram Analysis}
The proposed image encryption scheme was tested on different colour images yielding a flat histogram for the cipher image. For illustration, two different colour images are shown in Fig.~ \ref{Enc1} and \ref{Enc4}. One can see from these figures that the histogram of the encoded images is concentrated more toward the left side and hence it is evident that most of the pixels have values between 0 and 50. Furthermore, the pixel values are randomly distributed in the encrypted images. The histogram of the encrypted image shows that each pixel has approximately the same frequency of occurrence due to which the cipher images have flat histograms. 

\subsection{Statistical Analysis}
It can be observed from the results presented in Fig.~ \ref{Enc1} and Fig.~\ref{Enc4} that the encryption scheme conceals plaintext information. However, only visual inspection cannot guarantee the effectiveness of a good encryption technique. To obtain a quantitative assessment, Table \ref{tab:enc} exhibits the results after performing different statistical analyses on the two images shown in Fig. \ref{Enc1} and Fig. \ref{Enc4}. The values of the encryption parameters \cite{qayyum2020chaos} obtained in Table \ref{tab:enc} demonstrate the effectiveness of the presented DNA encryption technique. Detail explanation of the encryption evaluation parameters is given below: \\

\textbf{Correlation Coefficient:}
Correlation is a statistical technique for assessing the significance of the correlation between two variables. Correlation is a measure of how closely two variables are linked together. Any cryptosystem's correlation coefficient can be used to assess its level of encryption quality. An image cryptosystem will be considered strong if the encrypted image it produces is random and substantially uncorrelated, and the attributes of the input plaintext image are not present in the cipher image. The correlation coefficient between an encrypted image and the corresponding plaintext image must be exceptionally low if not zero. This means that if two images have a correlation coefficient equal to one, they would be considered identical. Correlation plots for the original and cipher images in different directions are shown in Fig. ~\ref{corr-image}. It is evident from Fig. ~\ref{corr-image} and Table \ref{tab:enc} that the correlation values are close to zero in all the directions. This implies that the pixels of the ciphertext images are uncorrelated. In contrast to the ciphertext correlation values, the original images correlation values are close to 1.  \\

\textbf{Entropy:} The entropy of a source provides insight into self-information or information that a random process provides about itself.  Entropy is a critical parameter to study when assessing an encryption method \cite{shannon1948mathematical}. Assume that we have a truly random source, $m = m_1, m_2...m_8$, which creates $2^{8}$ symbols having the same probability. The entropy value in this ideal scenario will be 8 bits. The entropy values calculated for the input and encrypted images are shown in Table \ref{tab:enc}. It is clear from Table \ref{tab:enc} that the entropy values of the ciphertext images are close to the ideal desired values. 
\\

\textbf{Key Sensitivity:} The key sensitivity test measures how sensitive an encrypted image is to a change in the key.  A reliable cryptosystem should decrypt the ciphertext image incorrectly if the key differs by only one bit. For extremely secure cryptosystems, large key sensitivity is necessary and changes in keys should generate a different ciphertext. This is done by calculating the difference between two ciphertext images when the encryption key for the two images is different by only one bit. Table \ref{tab:enc} shows the pixel difference between the plaintext and cipher images due to key sensitivity. It is evident from the results shown in Table \ref{tab:enc} that the proposed scheme has high key sensitivity, and therefore is sensitive to changes in the secret key.
\\

\textbf{NPCR and UACI:} The impact of a single-pixel change on the total image can be assessed using two standard methods: (i) NPCR and (ii) UACI. Generally speaking, the values of NPCR and UACI should be higher than 99\% and 33, respectively. It is evident from Table \ref{tab:enc} that the proposed encryption technique achieves adequately high values of the NPCR and UACI parameters.
\\

\textbf{Contrast:} Contrast analysis is a technique that analyzes an image's local intensity variation. An image's texture may be easily identified by a viewer using this statistical metric, which indicates texture uniformity. Varying gray levels can be seen in an image with higher contrast metrics, while images with lower contrast metrics show a lack of variation in gray levels. The contrast results shown in Table \ref{tab:enc} demonstrate the fact that the contrast values of the encrypted images are sufficiently higher than the original images. 
\\

\textbf{Homogeneity and Energy:} Homogeneity analysis is used to determine the closeness in the distribution of elements in the gray level co-occurrence matrix (GLCM). A low GLCM value of the encrypted image indicates a higher level of security. The energy parameter can be obtained by adding the squared values of the GLCM entries. A low energy value is obtained when the GLCM entries are similar, whereas a high energy value is obtained when the magnitude of some of the entries is greater than others. Secure images require low energy levels. From Table \ref{tab:enc}, it can be seen that both parameters have lower values thus depicting the effectiveness of the proposed image encryption scheme. 

\subsection{Comparison of the Keyspace}
The total possible combinations of keys that can be utilized in a cryptosystem's encryption and decryption are known as its keyspace size. The number of keyspace available can be used to estimate an encryption algorithm's strength. It has been observed that a minimum keyspace of $2^{100}$ is required to prevent a brute force attack. The computational precision, according to the IEEE standards, is roughly $10^{15}$, due to which there are $10^{285}$ keys for the proposed encryption technique, and these are significantly higher than $2^{100}$. Additionally, the approximate keyspace shown in Table \ref{tab:key} is higher than state-of-the-art encryption schemes.  \vspace{5mm}

%%%%%%%%%%%%%%%%%%%%%%%%%%%%%%%%%%%%%%%
\begin{table}[!ht]
\center
\renewcommand{\arraystretch}{1.8}
\caption{Keyspace analysis.}
\label{tab:key}       % Give a unique label
\scalebox{1.2}{
\begin{tabular}{llllll}
\hline
 & Ref \cite{wang2020hyperchaotic} & Ref \cite{sun2021image}   & Ref \cite{manikandan2022design} & Ref \cite{elkandoz2022image} & Proposed  \\
\hline
Keyspace &   $10^{75}$  &  $10^{96}$   & $10^{170}$ & $10^{168}$ &  $10^{285}$ \\  
\hline
\end{tabular}}
\end{table}

\FloatBarrier
\section{Conclusion}\label{sec6}
 To address the latency issue of encryption/decryption processes for large-size coloured images, a novel deep learning-based encryption scheme using DNA, chaos and multiple s-boxes is presented and evaluated in this paper. The proposed scheme consists of two modules; 1) a deep learning-based convolutional autoencoder to compress the large size three-dimensional coloured images into a significantly lower size two-dimensional gray-scale image, and 2) a novel DNA-based image encryption and decryption module using multiple chaotic sequences and substitution boxes to make the cipher image random as evident from the results reported in the paper. The convolutional autoencoder was trained on the Kodak image dataset with an impressive training and validation accuracy of 97\% and 95\%, respectively. The results of the autoencoder module demonstrated a high-quality reconstruction of the compressed images into the original colour images with an average PSNR of 29dB. Furthermore, in the proposed image encryption and decryption modules, four different chaotic maps; the TD-ERCS map, the Intertwining map, the Chirikov map, and the NCA map were utilized in conjunction with DNA coding and multiple s-boxes to effectively encrypt digital images. The security of the proposed encryption algorithm was evaluated using several parameters, such as histogram of the cipher image, entropy, NPCR, UACI, key sensitivity, contrast, etc. The results obtained after extensive experimentation revealed the effectiveness of the proposed convolutional autoencoder-based image encryption scheme in terms of; a) the successful compression of large-sized coloured images into low dimension grayscale images, b) the effective encryption/decryption of the compressed images and c) the high-quality reconstruction of compressed images into original three-dimensional colour images. The proposed scheme can effectively be utilized for secure and fast transmission of large-size coloured images over a low bandwidth transmission channel. A symmetric key technique is proposed in this paper for encrypting digital images. Symmetric encryption requires the same key for encryption and decryption. In future, we plan to transform the proposed scheme into an asymmetric image encryption scheme. Another interesting future direction could be to modify this scheme to work for video encryption.

\bibliographystyle{ACM-Reference-Format}
\bibliography{main.bib}
%\bibliography{references}
\appendix

\end{document}